\def\year{2022}\relax
\documentclass[letterpaper]{article} 
\usepackage{aaai22}  
\usepackage{times}  
\usepackage{helvet}  
\usepackage{courier}  
\usepackage[hyphens]{url}  
\usepackage{graphicx} 
\usepackage{natbib}  
\usepackage{caption} 
\DeclareCaptionStyle{ruled}{labelfont=normalfont,labelsep=colon,strut=off} 
\usepackage{amsmath}
\usepackage{graphicx}
\usepackage{subfig}
\usepackage{enumitem}
\usepackage{booktabs}
\usepackage{xargs}
\usepackage[pdftex,dvipsnames]{xcolor}
\usepackage{xspace}
\usepackage[nameinlink,capitalize]{cleveref}

\newcommand{\header}[1]{{\noindent{\textbf{#1}}}} 
\newcommand{\AITA}{{\textsc{AITA}}\xspace}
\newcommand{\aita}{\texttt{r/AmItheAsshole}\xspace}
\newcommand{\yta}{\texttt{YTA}\xspace}
\newcommand{\nta}{\texttt{NTA}\xspace}
\newcommand{\esh}{\texttt{ESH}\xspace}
\newcommand{\nah}{\texttt{NAH}\xspace}
\newcommand{\info}{\texttt{INFO}\xspace}
\newcommand{\NA}{\texttt{NA}\xspace}
\newcommand{\YA}{\texttt{YA}\xspace}
\newcommand{\topic}[1]{\textit{#1}}

\newcommand{\LW}{\texttt{LW}}
\newcommand{\NW}{\texttt{NW}}
\newcommand{\NE}{\texttt{NE}}
\newcommand{\KB}{\texttt{KB}}

\newcommand{\centercell}[1]{\multicolumn{1}{c}{#1}}
\usepackage{supertabular}
\usepackage{longtable}

\title{Mapping Topics in 100,000 Real-life Moral Dilemmas}

\author{
	Tuan Dung Nguyen, 
	Georgiana Lyall, 
	Alasdair Tran, 
	Minjeong Shin, \\
	Nicholas George Carroll, 
	Colin Klein, 
	Lexing Xie \\
}

\affiliations{
	Australian National University\\
	\{josh.nguyen, georgiana.lyall, alasdair.tran, minjeong.shin, nicholas.carroll, colin.klein, lexing.xie\}@anu.edu.au
}

\begin{document}
	
	\maketitle
	
	\begin{abstract}
		Moral dilemmas play an important role in theorizing both about ethical norms and moral psychology. Yet thought experiments borrowed from the philosophical literature often lack the nuances and complexity of real life. We leverage 100,000 threads --- the largest collection to date --- from  Reddit's \aita to examine the features of everyday moral dilemmas.  Combining topic modeling with evaluation from both expert and crowd-sourced workers, we discover 47 finer-grained, meaningful topics and group them into five meta-categories. We show that most dilemmas combine at least two topics, such as \topic{family} and \topic{money}. We also observe that the pattern of topic co-occurrence carries interesting information about the structure of everyday moral concerns: for example, the generation of moral dilemmas from nominally neutral topics, and interaction effects in which final verdicts do not line up with the moral concerns in the original stories in any simple way. Our analysis demonstrates the utility of a fine-grained data-driven approach to online moral dilemmas, and provides a valuable resource for researchers aiming to explore the intersection of practical and theoretical ethics.
	\end{abstract}
	
	\section{Introduction} \label{section:intro}
	\begin{figure*}[t]
		\centering
		\includegraphics[width=1\linewidth]{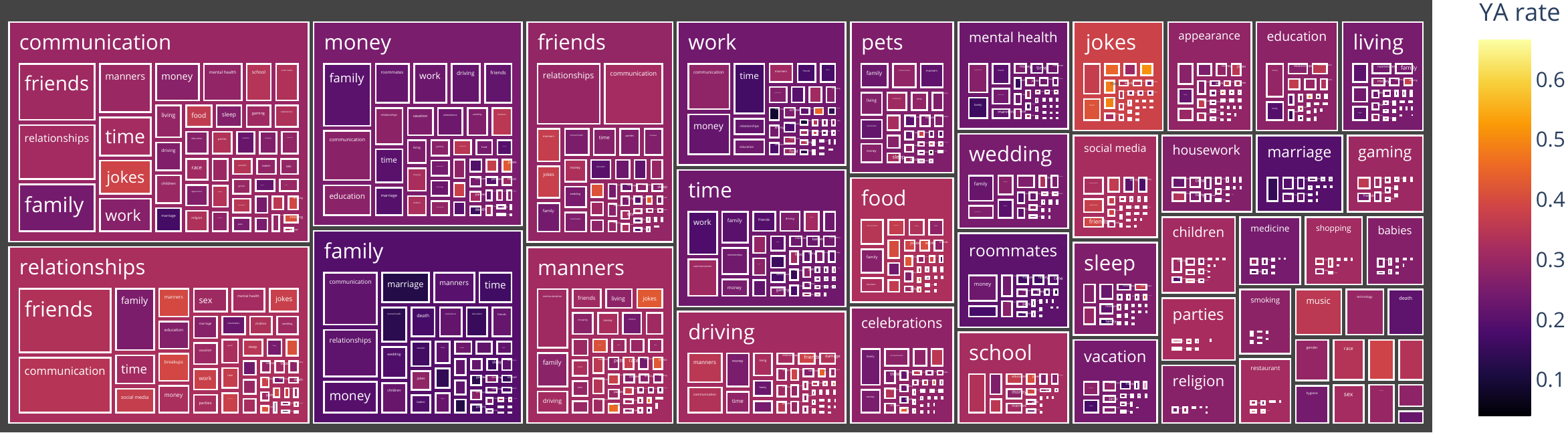}
		\caption{Treemap showing top-1 (outer rectangles) and top-2 (inner
			rectangles) LDA topics. The size of a block corresponds to the number of posts in a topic or a topic pair.
			{\em Communication} and {\em
				relationships} are the two most prevalent topics. 
			A lighter
			color corresponds to a higher proportion of \YA judgments ({\em you are the
				asshole} or {\em everyone sucks here}) within a
			topic/topic pair. For example, posts about {\em family}
			receives mostly a {\em not the asshole} judgment.}
		\label{fig:teaser}
	\end{figure*}
	
	Should we sacrifice the life of one person to save the lives of five others?
	Which patient should be prioritized in getting a kidney transplant? The idealized moral dilemmas that capture public imagination are clear and dramatic. This is by design. Thought experiments like the trolley problem \cite{ThomsonKilling76} make the conflict between moral principles especially stark. Daily life also presents people with a wide variety of comparatively small-scale, low-stakes, messy moral dilemmas. These remain under-studied because they lack the clarity of idealized dilemmas, yet they are arguably the sort of dilemmas that preoccupy most people most of the time.
	
	Philosophers define a moral dilemma as a situation in which an agent has a moral duty to perform two actions but can only perform one of them \cite{sinnott-armstrongMoralDilemmas1988}. Here we will use the term in a broader and non-traditional sense, encompassing {\em inter alia} what \citet{driverSuberogatory1992} calls a `morally charged situation'. This is a situation in which an agent is faced with a non-obvious choice between performing one of two actions, neither of which is morally required, but where one will elicit praise while the other will elicit blame. Such situations have attracted less attention, but  are equally important for understanding moral life. 
	
	In this work, we investigate {\em moral dilemmas that arise in daily life}. A broad study of such dilemmas will help to fill a crucial gap in philosophical and empirical inquiries into moral dilemmas. Few of us will allocate a kidney or sacrifice strangers; nearly everyone will have to deal with uncomfortable in-laws at a wedding, or adjudicate bitter debates over the workplace fridge. Many moral conflicts arise in pedestrian contexts from familiar concerns. A better understanding of everyday moral dilemmas will provide a novel foundation for testing philosophical and social scientific theories about the nature and taxonomy of our moral judgments such as the moral foundations theory \cite{grahamMappingMoralDomain2011}, morality as cooperation \cite{curryMoralityCooperationProblemCentred2016}, or forms of moral particularism \cite{dancy1983ethical,kagan1988additive}. Our analysis also shows that many moral dilemmas result from the interaction of what are traditionally considered conventional norms, suggesting that the moral/conventional distinction is less stark than some have supposed. Finally, a better understanding of everyday moral dilemmas could help shape the design of next-generation AI systems that are capable of fluid interaction in complex human environments.
	
	To this end, we turn to Reddit, a social network on which members can rate and discuss content submitted by other members. Reddit consists of user-created communities called {\em subreddits}, each of which focuses on a single topic of discussion. The \aita (\AITA) subreddit allows members to describe  a non-violent moral conflict that they have recently experienced, and ask the community to decide if they were in the right. \AITA is, as put by its community, {\em``a catharsis for the frustrated moral philosopher in all of us"}.\footnote{\url{https://reddit.com/r/AmITheAsshole}} It is a popular subreddit: at the time of writing, it has over 3 million members\footnote{A member of a subreddit is a user who subscribes to the subreddit. Active users who post or comment are typically a proper subset of all members.} and regularly ranks in the top 10 for volume of comments per day.\footnote{According to \url{https://subredditstats.com}} This makes it an excellent source for real-life moral dilemmas and the discussion that surrounds them.
	
	We extract more than 100,000 real-life moral dilemmas on \AITA, and design a multi-stage topic
	discovery process using both expert and crowd-sourced validation to map 94\% of these scenarios into 47 interpretable topics. We posit that topics are an informative lens through which to study \AITA. Our rigorous discovery and validation process is designed to eliminate ambiguities that will propagate to subsequent analysis. These topics need not be mutually exclusive, and indeed the richness of content of \AITA dilemmas means that most are better characterized by a topic pair instead of a single topic. As \cref{fig:teaser} shows, \AITA topic pairs vary both in popularity and in the judgments they attract. Many \AITA dilemmas involve traditionally non-moral domains, suggesting a more nuanced structure than those of traditional philosophical thought experiments.
	
	The main contributions of this work include:
	\begin{itemize}[leftmargin=0.7cm, noitemsep]
		\item Curating a large collection of everyday
		moral dilemmas, which is publicly released\footnote{The dataset and code can be found at \url{https://github.com/joshnguyen99/moral_dilemma_topics}};
		\item A novel data-driven topic discovery method with multiple stages of validation to map these dilemmas into five meta-categories spanning 47 meaningful topics; 
		\item Demonstrating ways that an understanding of everyday moral dilemmas can produce new insights into philosophical discussions relating to moral theorizing; and
		\item Empirical insights showing how everyday moral dilemmas are generated by combinations of topic pairs, how certain topics attract or repel other topics,
		and how the moral valence of similar words can vary across different topic	pairs.
	\end{itemize}
	
	\section{Related Work} \label{section:related_work}
	This work is related to the rich literature on moral dilemmas, topic modeling and discovery, and online collective judgment and decision making.
	
	\subsubsection{Moral dilemmas.}
	Moral dilemmas \cite{sinnott-armstrongMoralDilemmas1988} and morally charged situations \cite{driverSuberogatory1992} play a crucial role in philosophical theorizing. There is an empirical literature aimed at teasing out the mechanisms that drive individual judgments about classic dilemmas \cite{GreeneAn-fMRI01};  this work has become increasingly important in informing moral domains like algorithmic decision-making systems such as driverless cars \citep{AwadThe-moral18} or kidney exchange programs \citep{FreedmanAdapting20}. 
	
	We note three features that characterize much of the existing empirical work and set it apart from the current study.  First, existing work tends to focus on  stark dilemmas -- like the so-called `trolley problems' \cite{ThomsonKilling76} -- that require individuals to pass judgment on unfamiliar and unrealistic situations. Second, existing work tends to rely on survey data or laboratory experiments rather than conversations with peers. Moral judgment and justification are sensitive to perceived beliefs and intentions of one's audience, including the experimenters themselves \cite{TetlockAccountability83}. Hence such settings may not reveal the full range of the participants' reasoning. Observational posts of online social media  represent the sort of `unobtrusive measure' \cite{WebbUnobtrusive99} that can avoid experimenter effects. Third, existing work tends to give subjects pre-packaged, simple moral dilemmas. Yet figuring out how to frame a moral problem in the first place is often an important issue in its own right \cite{AppiahExperiments08}. By contrast, \AITA represents a rich source of moral dilemmas that are realistic and familiar, presented by an involved party as part of a conversation with peers, and in a forum that allows for dynamic probing and re-framing the issues at hand. The \AITA dataset thus represents a valuable resource for studying moral dilemmas and crowdsourced judgments, one that can compliment existing hypothesis-driven work.
	
	\subsubsection{Topic modeling in text.}
	The task of understanding large document collections is  sometimes referred to as {`describing the haystack'}. Data clustering approaches are widely used for such problems.  Methods that are specifically designed for text data include Probabilistic Latent Semantic Indexing \cite[pLSI,][]{hofmannProbabilisticLatentSemantic1999} and Latent Dirichlet Allocation \cite[LDA,][]{bleiLatentDirichletAllocation2003}.  In particular, LDA has been widely applied to historical documents, scientific literature, and social media collections \cite{boydgraberApplicationsTopicModels2017}, to name a few. 
	
	We categorize the evaluation of topic models into intrinsic and extrinsic methods. Intrinsic methods evaluate components of the topic models themselves. Held-out data likelihood \cite{bleiLatentDirichletAllocation2003} has been the \textit{de facto} choice when evaluating an entire LDA model intrinsically. For each topic, human-in-the-loop approaches with intruding words \cite{changReadingTeaLeaves2009}, or coherence metrics based on the probability of word co-occurrences \cite{mimnoOptimizingSemanticCoherence2011} have been shown to correlate with human judgments. Extrinsic methods evaluate topic models with respect to domain-specific tasks; examples of these are as diverse as the application domains. In the scientific literature, topic model outputs have been compared against surrogate ground truths such as author-assigned subject headings, and used for trend spotting over time \cite{griffithsFindingScientificTopics2004}. In analyzing historical newspapers, \citet{newman2006probabilistic} annotated a subset of topics of interest to history and journalism but were not concerned with either covering the whole dataset or ensuring most topics are meaningful. In literature, derived statistics from topics have been shown to evaluate specific conjectures about gender, anonymity, and literary themes \cite{jockers2013significant}. Recently \citet{antoniak2019narrative} used LDA to discover narrative paths and negotiation of power in birth stories.  Their dataset is much smaller (2.8K) and more topically concentrated than the  \AITA dataset used in this paper. Also, their topics are validated using an existing medical taxonomy, whereas there is no such resource for everyday moral conflicts.
	
	What differentiate this work in the application of topic models are the striving for coverage of a large collection, and the goal of supporting both qualitative and quantitative tasks. To the best of our knowledge, the two-stage validation combining the opinions of experts and general users is new.
	
	\subsubsection{Moral judgments on social media.}  This topic area is quickly gaining momentum in computational social science.  Two recent works are focused on analyzing language use in moral discussions. \citet{zhouAssessingCognitiveLinguistic2021} profiled linguistic patterns in relation to moral judgments, showing that the use of first-person passive voice in a post correlates with receiving a not-at-fault judgment. \citet{haworthClassifyingReasonabilityRetellings2021} called the judgment on a post `reasonability' and built machine learning classifiers to predict the judgments using linguistic and behavioral features of a post.  Other works are focused on automated prediction of moral judgements. \citet{botzerAnalysisMoralJudgement2021} built a moral valence (\yta and \nta) classifier on \AITA data and evaluated its utility on other relevant subreddits.  Delphi \cite{jiangDelphiMachineEthics2021} is a research prototype that takes in a one-line natural language snippet and gives a moral judgment from a wider range of possibilities (e.g., {\em expected, understandable, wrong, bad, rude, disgusting}). Its large-scale neural model is trained on multiple data sources including parts of \AITA. The related Social Chemistry project \cite{forbes2020social} breaks down judgments of one-liner scenarios into rules of thumb, covering social judgments of good and bad, moral foundations, expected cultural pressure and assumed legality.
	
	While recent work focuses on directly correlating the natural language content (of a post, a title snippet, or a comment) with moral judgments, we choose to focus on taxonomizing the structure of moral discussions as a first step. We posit that there are diverse practices used by the online community in moral argument and reaching a verdict as a group. This hypothesis is supported in \Cref{section:observations}, showing that topics are an important covariate for the differences in the moral foundation to which posters appeal.
	
	\section{Dataset} \label{section:dataset}
	
	\subsection{Structure of \aita}
	In a subreddit, discussions are organized into {\em threads}. Each thread starts with a {\em post}, followed by comments. Each post consists of a {\em title, author, posting time}, and {\em content}; and each comment contains an {\em author, timestamp, content}, and {\em reply-to} (the ID of a post or another comment). Community rules dictate that a post title must begin with the acronym `AITA' or `WIBTA' (Would I Be The Asshole?).
	
	Collective judgments are reached via {\em tagging} and {\em voting}. Five types of judgments are defined in \AITA: \yta (you are the asshole), \nta (not the asshole), \esh (everyone sucks here), \nah (no asshole here), and \info (more information needed). Each comment can contain one of these tags. A user can cast an upvote (scoring +1) or a downvote (scoring -1) to a comment. The judgment of the top-scoring comment would become the community verdict, called {\em flair}, and be displayed as a tag for the post.\footnote{\url{https://mods.reddithelp.com/hc/en-us/articles/360010513191-Post-Flair}} The {\em flair} of a post is assigned by a bot after 18 hours.\footnote{This timeframe was chosen by the community. The full process is documented in the \AITA community FAQ \url{https://www.reddit.com/r/AmItheAsshole/wiki/faq}.} Appendix Fig. A1 shows an example thread with the \yta flair, and another comment judging it as \nta.
	
	\subsection{The \AITA dataset} \label{ssec:data_AITA_posts}
	
	We use the Pushshift API \cite{baumgartnerPushshiftRedditDataset2020} to retrieve  all posts and comments on \AITA from 8 June 2013 to 30 April 2020, yielding 148,691 posts and 18,533,347 comments. When a post's flair maps to a judgment (such as \nta), we use it as the post's verdict. In the 946 posts without a valid flair, we reconstruct each post's verdict using the judgment contained in its highest-scoring comment. After this, 920 posts remain without flairs. To filter out moderation and meta posts, we keep posts whose titles start with `AITA' or `WIBTA', have at least 50 words, 10 comments, 1 vote, and 1 verdict. This yields 108,307 posts and 8,953,172 comments. Posts with fewer than 10 comments consist only of 20\% of the dataset and are generally of lower quality. We use the 102,998 threads in or before 2019 as our training set, and 5,309 threads in the first four months of 2020 ,as the test set for the topics discovered (\cref{section:topic_evaluation}). When pair-wise comparison is called for, we group \nta and \nah into the \NA judgment class with \emph{positive valance} on the original poster. Similarly \yta and \esh are grouped into the \YA class with \emph{negative valance}.
	
	We note that works using the Pushshift Reddit API that were published before 2018 may have involved missing data, which can lead to systematic biases in downstream analyses \cite{gaffneyCaveatEmptorComputational2018}. However, \citet{baumgartnerPushshiftRedditDataset2020} have since recrawled the missing posts and thus our derived data is less likely to suffer from the same problem.
	
	\begin{figure}
		\centering
		\subfloat[]{%
			\includegraphics[height=3cm]{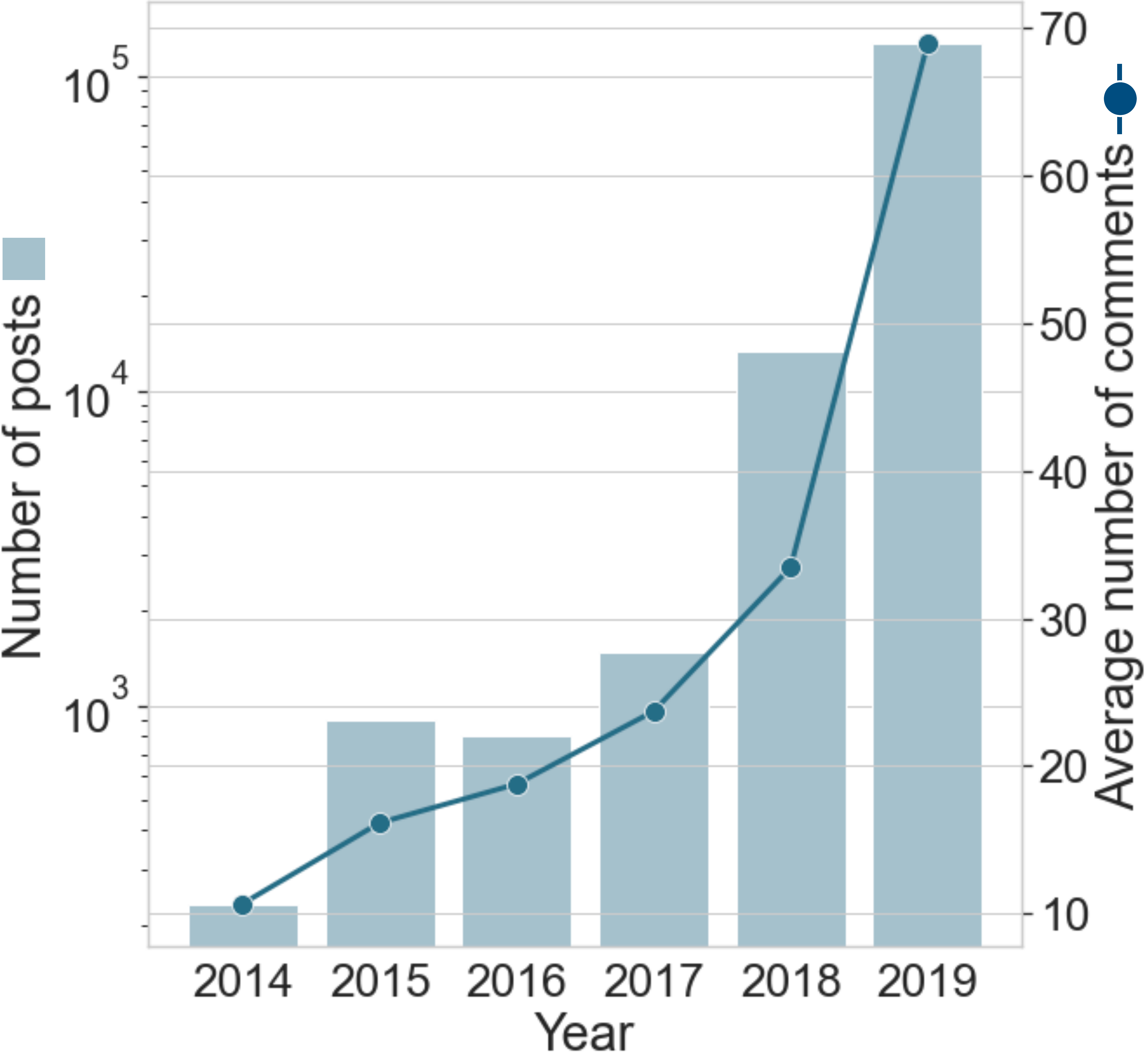}%
			\label{fig:dataset:traffic:posts}%
		}
		\
		\subfloat[]{%
			\includegraphics[height=3cm]{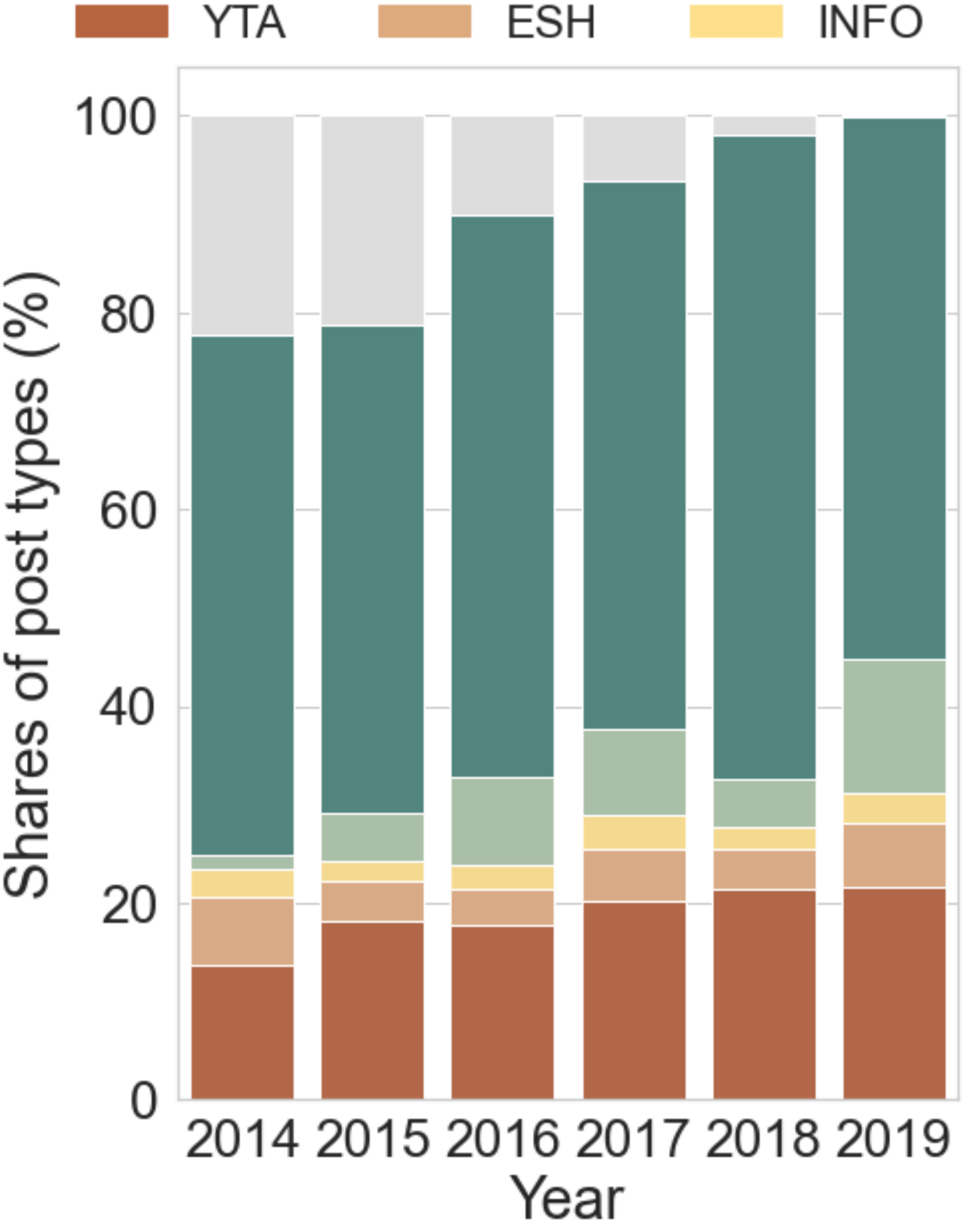}%
			\label{fig:dataset:label_share}%
		}
		\
		\subfloat[]{%
			\includegraphics[height=3cm]{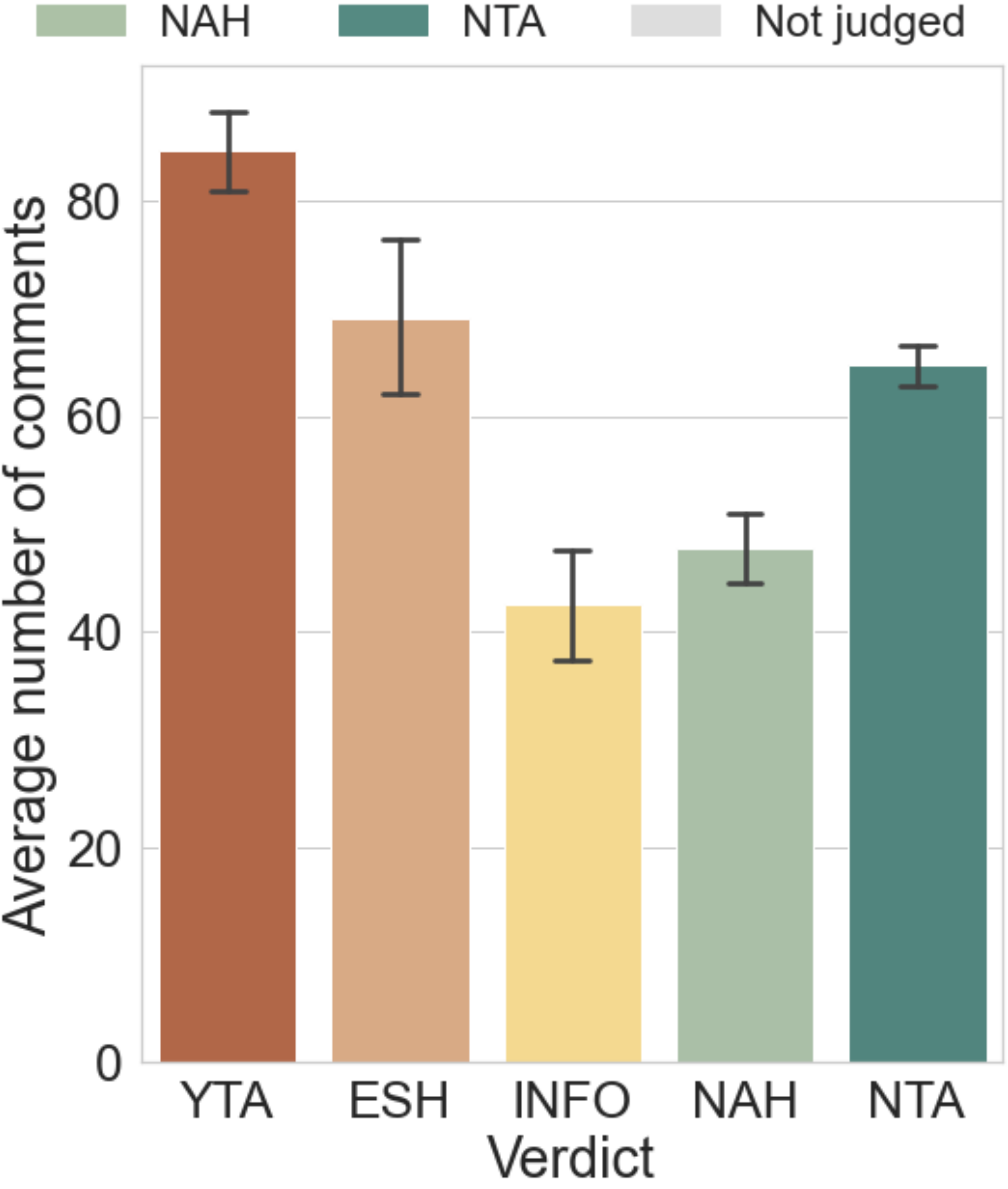}%
			\label{fig:dataset:avg_comments}
		}
		\caption{Data statistics of \AITA. \protect\subref{fig:dataset:traffic:posts} Number of posts and the average	number of comments 2014--2019. \protect\subref{fig:dataset:label_share} Shares of judgments over time. \protect\subref{fig:dataset:avg_comments} Average number of comments (and 95\% CI) for posts with each judgment.}
		\label{fig:dataset:traffic}
	\end{figure}
	
	\cref{fig:dataset:traffic} presents the number of posts, number of comments per post and breakdown of flairs. Over time, participation increased quickly as more members entered the subreddit. Both the number of posts and the average number of comments per post rose over the years, with 2018 and 2019 seeing the most significant increases (note the y-axis in log scale). The flair shares remained consistent in 2018 and 2019, with \nta posts taking more than half of the posts (65.32\% in 2018 and 55.14\% in 2019). In terms of controversiality, negatively judged posts (with flair \yta or \esh) tend to attract more comments than positively judged posts (with flair \nta or \nah). When looking at the post lengths (shown in Appendix Fig. A2), \esh posts are the longest on average (mean = 433.2 words), reflecting the nuances required when describing situations with no clear winner. We also observe that \nta posts tend to be longer than \yta posts (\nta: mean = 400.6; \yta: mean = 370.6), while \yta posts attract more comments overall (\nta: mean = 79.4; \yta: mean = 107.6).
	
	\section{Discovering topics on \AITA}
	\label{section:topic_modeling}
	
	We adopt a data-driven topic discovery process with two stages of manual validation, as outlined in \cref{fig:flowchart}. An exploratory study that shaped our clustering choices is described in Appendix Section B. Taking as input the 102,998 posts until the end of 2019 as the training set, we use text clustering algorithms to group the collection into clusters and describe their properties.
	
	Clustering methods discover self-similar groups in data, called clusters. Given the goal of mapping different kinds of moral dilemmas on \AITA, the ideal set of clusters should have a high {\em coverage} of the whole dataset, and  the clusters (and posts within) should be {\em distinguishable} from each other as judged by human readers. Our choices of which clustering methods to use are informed by the desiderata from the work of \citet{vonluxburgClusteringScienceArt2012}. Firstly, our task is {\em exploratory} rather than confirmatory. Secondly, the use of the resulting clusters is both {\em qualitative} (in grounding the types of dilemmas to moral philosophy) and {\em quantitative} (for measuring behavioral and linguistic patterns of the resulting clusters). Moreover, we prefer clustering algorithms that allow clusters to overlap, since both the intersections and the gaps between two intuitive clusters (such as \topic{family} and \topic{money}) may be meaningful and interesting.
	
	The rest of this section discusses the choices and trade-offs made in
	clustering posts (\Cref{ssec:clustering}), the manual validation that turns
	clusters into named topics (\Cref{ssec:naming}), and observations of the
	resulting topics (\Cref{ssec:named_topics}).
	
	\begin{figure}
		\centering
		\includegraphics[width=0.98\linewidth]{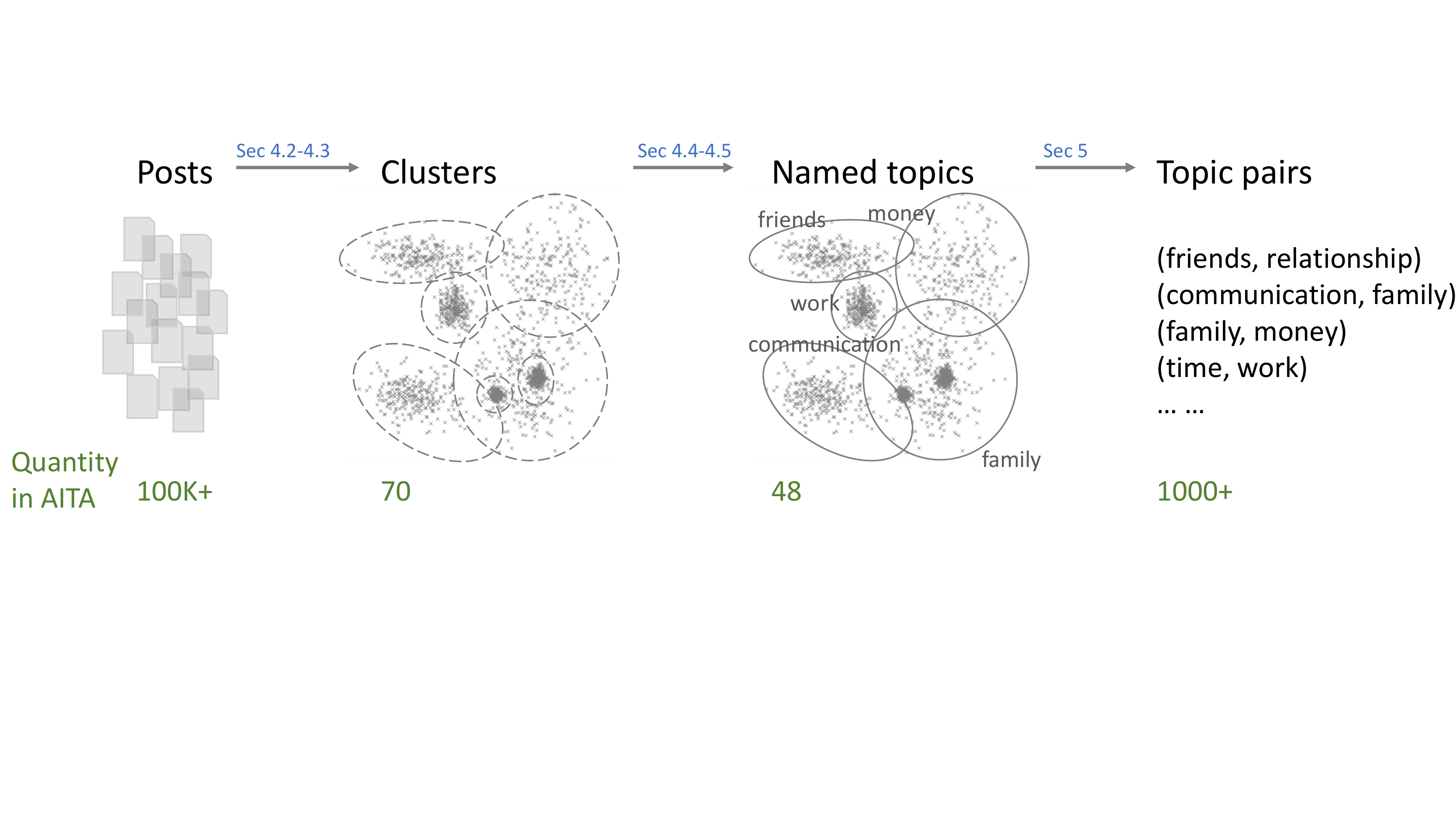}%
		\caption{A high-level overview of the discovery process of AITA topics, with two stages of human validation indicated by {\bf *}. Quantities at the bottom indicate the size (number of posts, clusters, topics and topic pairs) after each stage.}
		\label{fig:flowchart}
	\end{figure}
	
	\subsection{Clustering posts} \label{ssec:clustering}
	
	We perform probabilistic clustering using LDA. The input to LDA is a set of vectors containing word counts for each post. To create these vectors, we tokenize a post's body (excluding its title), lemmatize each token, remove stop words, and eliminate tokens which appear in fewer than 20 posts, all using \texttt{spacy} \cite{honnibalSpaCyIndustrialstrengthNatural2020} and \texttt{scikit-learn} \cite{Pedregosa2011Sklearn}. We keep $M=10,463$ words across all training posts and denote these words as $x_m$, $m=1,\ldots,M$. The outputs from LDA are two sets of probabilities. First, the representation for each of the $K$ clusters $k \in \{1,\ldots,K\}$ is a multinomial word probability vector $p(x_m \mid k)$. This probability is sorted to produce the top words for each topic, which  helps interpret the clusters.  Second, the posterior probability of each cluster given each document $d$ is $p(k \mid d)$, representing the salience of each topic within a document. They are sorted to produce the top cluster(s) for each document. Both probabilities will be used in topic evaluation and interpretation (\Cref{ssec:naming,section:topic_evaluation}). While one main limitation of LDA is the use of unordered bag-of-words representations, the two probability representations lend themselves to direct human interpretation of the topics, which ensures that topics are {\em distinguishable} from each other. Moreover, representations generated from LDA support {\em overlapping} topics, both {\em qualitative} and {\em quantitative} analysis of topics, and discovery of trends and behavioral patterns.
	
	\header{Choosing the number of clusters} is an important practical question for topic discovery, and greatly affects the {\em coverage} and {\em distinguishability} of the resulting clusters. We first examine the document perplexity on a held-out dataset (Appendix Section C.1), which indicates that the optimal is around 40 clusters (Appendix Fig. C1). However, upon examining the sizes of the resulting clusters (by assigning documents to their top-scored cluster), we find that several clusters are too big in size ($>15\%$ of the dataset) and appear uninformative by their top keywords and top documents. We therefore increase the number of clusters to 70, which results in more balanced clusters ranging from 0.02\% to 7.63\% in size, all of which go through a subsequent vetting process by human experts (\cref{ssec:naming}), resulting in 47 named topics after merging and pruning clusters. Note that it is not possible to set the number of clusters equal to 47 \textit{a priori}, since clustering algorithms are influenced by random initialization and prone to producing a few clusters that are similar to each other \cite{boydgraberApplicationsTopicModels2017}.
	
	\subsubsection{Alternatives in text representation and clustering.} Besides LDA on bag-of-words, we experiment with other models such as non-negative matrix factorization \cite{paateroPositiveMatrixFactorization1994} and soft K-means \cite{dunnFuzzyRelativeISODATA1973} and with other embedding methods such as TF-IDF (10,463 dimensions), Empath \cite[194 dimensions]{fastEmpathUnderstandingTopic2016} and Sentence-RoBERTa \cite[1,024 dimensions]{reimersSentenceBERTSentenceEmbeddings2019}. While each method has its merits, we find that LDA described in this section is the most suitable. Detailed description and comparisons can be found in Appendix Sections C.2 -- C.4.
	
	\subsection{Cluster evaluation overview}
	\label{ssec:whyeval}
	
	LDA topics, just like outputs from other clustering algorithms, 
	contain several sources of ambiguity and noise that make them difficult to use for downstream interpretation or moral reasoning tasks. 
	First, clusters are defined by patterns of co-occurrence in data, but categories of stories need semantically recognizable {\em names} in order to support moral reasoning and generalization. Second, the correspondence between clusters and names is rarely one-to-one: there are often semantically similar clusters that share a name, or meaningless clusters defined by functional words for a domain, such as {\em edit}, {\em upvote}, {\em OP} (original poster) for Reddit. Such noise is well-known in practice, and a body of work has been devoted to topic model evaluation, stability and repair \cite[Section 3.4]{boydgraberApplicationsTopicModels2017}. 
	
	We design a rigorous two-stage evaluation for moral topics. 
	The first stage is {\em naming} topics, covered in Section \ref{ssec:naming}.
	This is driven by the need to having {name} topics in ways that are more succinct, semantically comprehensible, and free of the above noise. 
	This process is called labeling in the topic model literature \cite{boydgraberApplicationsTopicModels2017}. We opt to name topics manually, rather than automatically, which will not be able to prune meaningless clusters. 
	Topic naming is done by a small number of {\em experts} (co-authors of this paper, including both philosophers and computer scientists) 
	because they need to be familiar with the LDA internal representation of ranked list of words, and also because of the need to deliberate (described in \cref{ssec:naming}) when names are semantically similar but not identical.
	
	The second stage is intended to validate the utility of the assigned names to a broad audience of online crowd workers.   
	This is to ensure that the named topics are widely recognizable, and that the names are appropriate for the posts in the corresponding clusters. See Section 5 for details.
	
	\subsection{From clusters to named topics} \label{ssec:naming}
	
	The unit for this annotation task is a cluster $k$, ($k\in
	\{1,\ldots,70\}$). A screenshot of this web-based survey is shown
	in Appendix Fig. D1. Each question starts with macroscopic
	information about the cluster -- the top-10 keywords sorted by word probability
	$p(x \mid k)$. Showing 10 words is a common practice in LDA
	evaluation \cite{newmanAutomaticEvaluationTopic2010}. This is followed by a
	microscopic view of the cluster -- the content of three top posts, sorted by
	posterior probability $p(k \mid d)$, and three randomly chosen posts whose top-scoring cluster
	is $k$. By default, the list of posts is shown with the titles only, which can
	be expanded to show the first 100 words of the post by clicking on the title. For each task, the annotator is asked to provide a name for the cluster consisting of one or two words, or to indicate that a coherent name is not possible with {\em N/A}.
	
	Six authors of this paper participated in this annotation task. We collect three independent answers per question from three different annotators. Anonymized inputs are collated in a spreadsheet. Two of these annotators are then designated to resolve disagreements in naming. They review the results and make four types of decisions to name the 70 clusters: \textbf{unanimous}, \textbf{wording}, \textbf{deliberation} and \textbf{other}. There are 17 clusters with \textbf{unanimous} agreement, in which all three annotators agree on the exact wording, e.g., {\em shopping} and {\em pets}. Meanwhile, in 41 clusters, the names for the same cluster have very similar semantic meaning but exhibit \textbf{wording} variations such as synonyms. In this case, one of them is chosen based on brevity and specificity, e.g., {\em race} was chosen over {\em racism} and {\em babies} over {\em pregnancy}. A \textbf{deliberation} between the annotators is required for 9 clusters where different names are present. Here the annotators take into account whether there are two inputs that agree, the semantics of the top words, and the distinctiveness from other topics. For example, three annotators assign (\topic{appearance}, \topic{tatoos}, \topic{appearance}) to a topic, and {\em appearance} is chosen after re-examining the keyword list and discussing the scope of the topic. Finally, there are 3 clusters with no agreement even after discussion. These are grouped into a placeholder topic \textbf{other}. Clusters with the same name are merged: 67 clusters are merged into 47 named topics in this process, with topic \topic{family} having the most repetitions of 5. After merging, we end up with {\em 47 named topics} (96,263 posts or 93.5\%) and a placeholder topic \topic{other} (6,735 posts or 6.5\%). The topic \topic{other} will be excluded from subsequent sections. See Appendix Section D for more detail. Finally, as some topics are merged from several clusters, we aggregate the posteriors of clusters $c$ with the same name into a \emph{topic posterior} for topic $k$:
	\begin{align*}
		p(k \mid d) = \sum_{c: ~\text{name of} ~c = k } p(c \mid d).
	\end{align*}
	Throughout the rest of the paper, we refer to this definition when talking about the topic posterior. For example, the top-1 topic given document $d$ is ${\mathrm{argmax}}_k ~p(k \mid d)$.
	
	This cluster annotation task is conducted by human experts as it requires an
	understanding of the \AITA domain, the goal of topic mapping, and a
	high-level knowledge of what LDA keywords represent. The deliberation cannot easily be done in an online distributed setting. The apparently
	low fraction of unanimous agreement in this free-form naming task is
	consistent with what we observe in a topic discovery exploration (Appendix Section B). The named topics are then validated using
	crowd-sourcing by evaluating the match between topic names and post content in
	\Cref{section:topic_evaluation}.
	
	\begin{figure}
		\centering
		\includegraphics[width=1\linewidth]{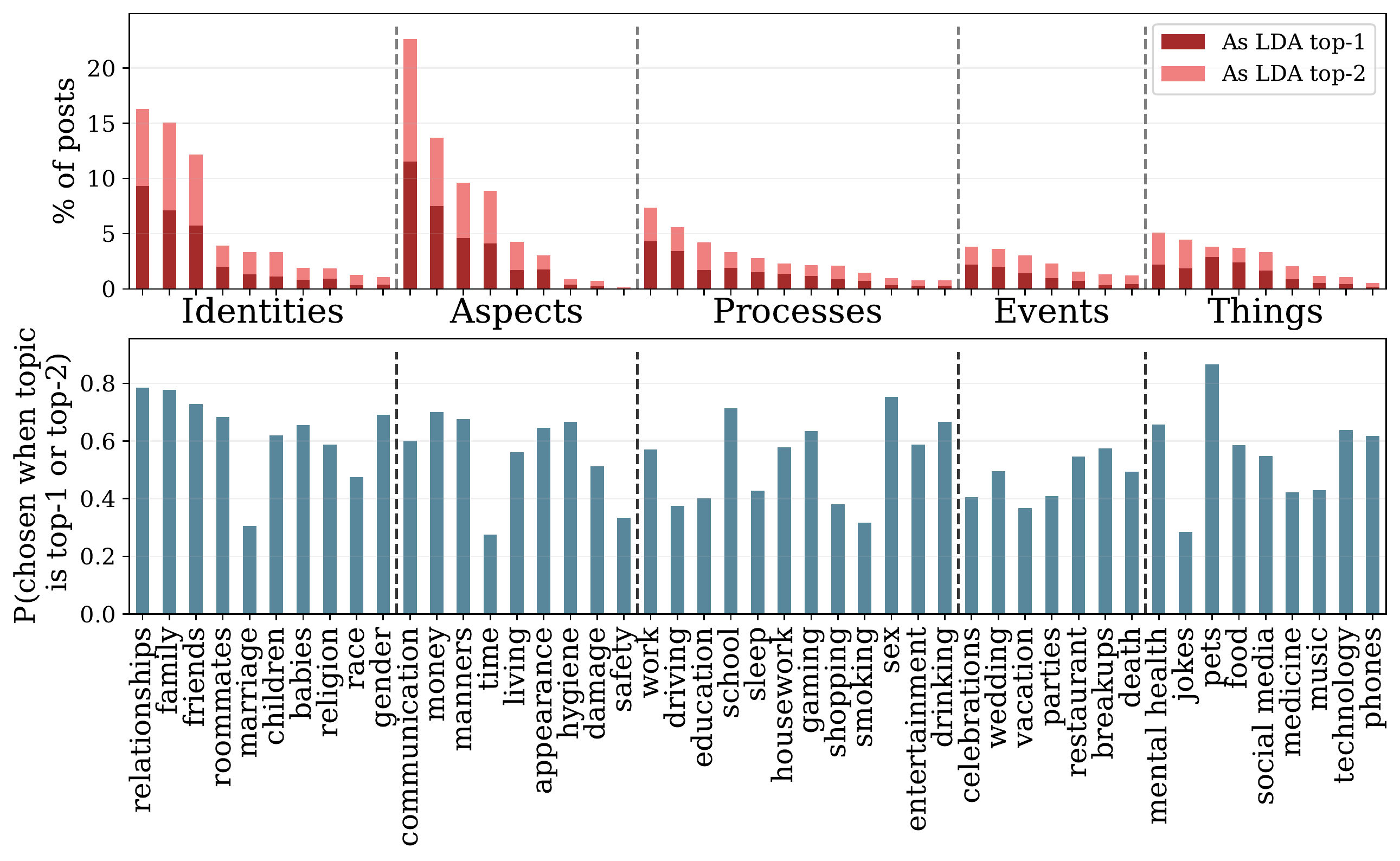}
		\caption{Topic level statistics on the {\em training} set, grouped by their categories. The top figure shows the prevalence (as a percentage) of each topic as the top 1 or top 2 for all post (see \cref{section:topic_modeling}).
			The bottom figure shows the topic-specific agreement rate (see \cref{section:topic_evaluation}).}
		\label{fig:topic_stats_train}
	\end{figure}

	\subsection{A summary of named topics}
	\label{ssec:named_topics}
	
	As an aid to navigate the set of topics, we further group the 47 named topics (less \topic{other}) into five meta-categories. \emph{Identities} (individuals and their social relationships to others) and \emph{things} (other themes)  broadly correspond to static narrative roles. Topics with a dynamic aspect are grouped into \emph{processes} (things that happen indefinitely or regularly), \emph{events} (specific one-off occasions that are individually important), and \emph{aspects} (the manner in which a process or event occurs). The meta-categories, chosen by author consensus, are meant as a heuristic aid to interpretation; other carvings are possible, assignments might vary, and individual topics might cross boundaries. For example, we group \topic{sex} as a process because many \AITA posts are about the poster's sex life, which is an indefinite ongoing process, but individual instances of sex might be better considered as events. Nevertheless, our rough grouping of topics  aids  interpretability. The list of topics along with their frequencies are shown in \cref{fig:topic_stats_train} (top), grouped by meta-categories and sorted by their prevalence within. We can see that the most frequent topics are all within {\em identities} and {\em aspects}, likely due to the fact that \AITA posts are often generated by social conflicts defined by relations to and manner of interactions with others.
	
	We have five observations on the topic list. The first is that common scenarios in one's social life are covered -- from family to professional relationships, from work to recreation. The second is that the topics are neither  exhaustive nor fine-grained. For example, there is no topic on medical moral dilemmas common in TV dramas, likely due to their rarity in daily life. Some intuitive `topics' are absent but get coverage by their individual aspects. For example, there is no \topic{travel} topic, but there are topics covering \topic{vacation}, \topic{work}, \topic{money} and other individual aspects of travel. The third is that the prevalence of posts classified under topics such as \topic{communication} and \topic{manners} suggest that the way in which an action is performed is presented as morally salient. The fourth is that the relative prevalence of topics can change over time. Comparing to a validation set of 982 posts on the last three days of 2019, \topic{family} and \topic{celebrations} rose significantly, whereas \topic{school} and \topic{driving} dropped. Finally, it is surprising that the posterior probability of the top-ranked topic for each post tends to be fairly close to that of the second-ranked topic (mean difference 0.141, see Appendix Fig. D2 for examples). This suggests that the top few topics for each post may be similarly relevant, rather than only the top topic being significantly relevant to a post.
	
	\section{Crowd-sourced topic survey} \label{section:topic_evaluation}
	
	We design and conduct a set of crowd-sourced surveys to answer two key questions: how well do human annotators agree with the named topics, and how do users at large perceive topics of an \AITA post? A complete description of the survey is found in Appendix Section E.
	
	\subsection{Survey setup} \label{ssec:eval_setting}
	
	\begin{figure}[!t]
		\centering
		\includegraphics[width=1\linewidth]{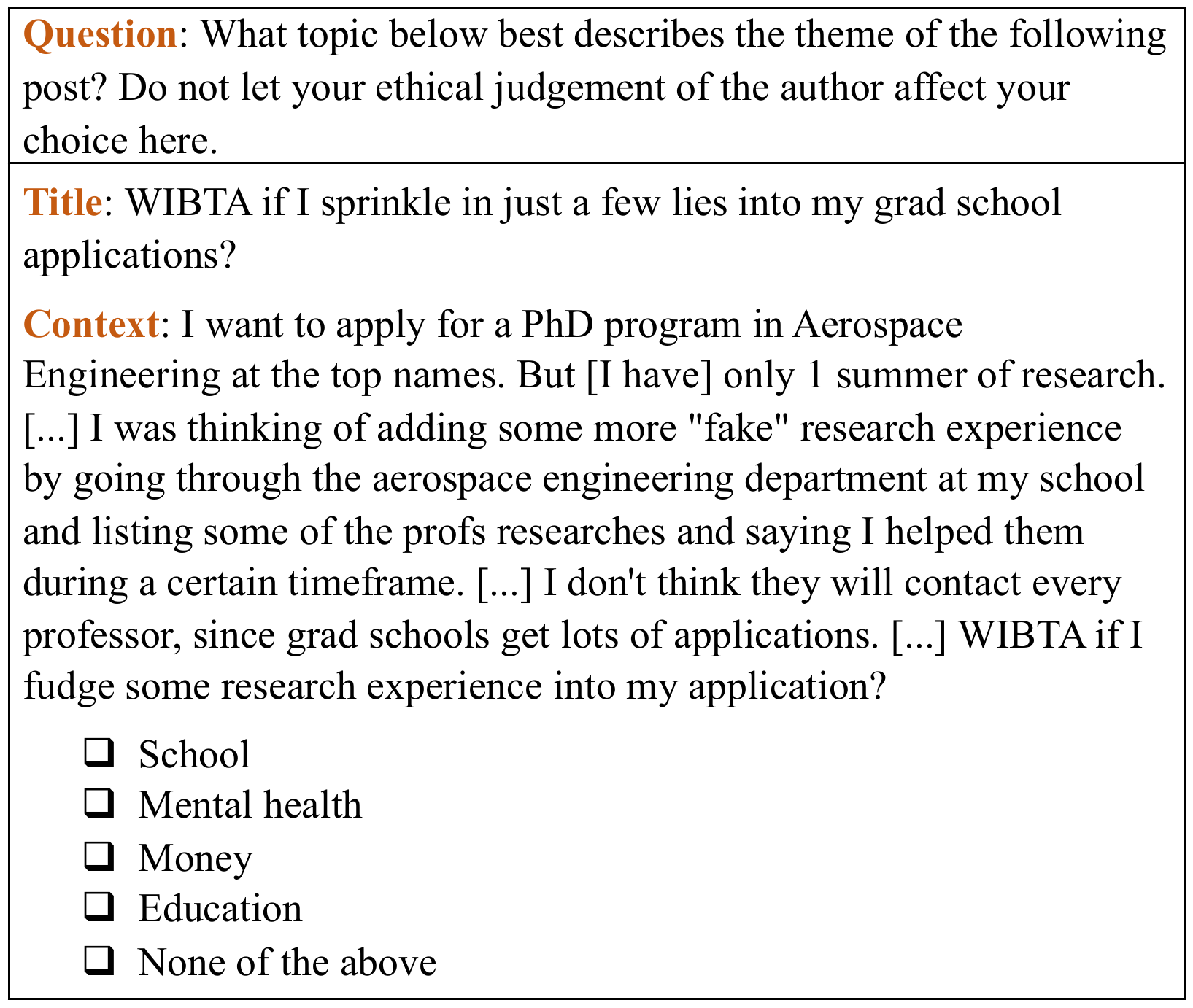}
		\caption{One example question in the topic validation survey. Each question
			contains a post (title and body) and has four topic options. The
			participant can choose more than one option, or \textit{None of the above} if no topic name matches the post. The top 4 topics according to
			the LDA posterior of this post are \topic{education}, \topic{school},
			\topic{money}, and \topic{mental health}.}
		\label{fig:crowdsourcing:post_to_topic}
	\end{figure}
	
	Each crowd-sourced survey consists of a number of questions, each of which is centered on an \AITA post and starts with a fixed prompt: {\em ``What topics below best describe the theme of the following post? Do not let your ethical judgement of the author affect your choices here.''} We then present the post title and body text, and five topic choices. The first four choices are a randomized list of the top 4 topics according to the topic posterior, followed by a \emph{None of the above} option. A participant can choose one or more non-conflicting options before moving on to the next post. An example question is shown in \cref{fig:crowdsourcing:post_to_topic}. Free-form text boxes are also provided to collect participants' reflections at the end of each question, as well as at the end of the survey.
	
	We use the Prolific crowd-sourcing platform\footnote{\url{https://www.prolific.co}} to recruit participants. Each individual can only enter once, and we collect answers from three different participants for each question. To control the quality of results, we only allow fluent English speakers to participate. Before entering the actual survey, each participant is given one training question, containing one post clearly belonging to two of the given topics. Choosing the correct answers for this training question is a prerequisite for completing the rest of the survey. 
	
	Based on a pilot test among the authors, we set the length of each survey to 20 questions, with a time estimate between 12--20 minutes. A total of 285 participants (130 males, 151 females, 4 unspecified) completed the survey, each of whom was paid \textsterling 2.5 for their work. Their average age is 28.2 (SD=9.2), with 39.1\% living in either the US or UK. This survey design is approved by the ANU Human Research Ethics Committe (Protocol 2021/296). More information about this survey and participation statistics can be found in Appendix Section E.4.
	
	We collect survey results in three settings. On the {\em training} split of \AITA (\cref{section:dataset}), we randomly select $20$ posts for each topic, and call this setting \textbf{train}. The topic choices are the top $4$ choices according to the LDA posteriors. We increase the size of the survey with 5, 10 and 20 posts per topic gradually, and find that the statistics stabilize after 10 posts per topic. On the {\em test} split of \AITA, which is not seen by either the LDA estimation or in topic naming, we randomly select 10 posts for each topic populated with its top 4 topics, and call this setting \textbf{test}. This gives us 450 posts in total. Note that for 5 topics with fewer than 10 posts, we simply include all the posts. Lastly, we use the same set of posts from the \emph{test} set, but include the top-2 topics according to LDA, plus two other randomly selected distractor topics for each post. We call this setting \textbf{test+rand}, which is designed to  observe whether or not the top 2 topics are significantly more descriptive than other randomly selected topics. These three settings are shown as column headings in \cref{table:post_to_topic_results}.
	
	\begin{table}
		\centering
		\caption{Post-level agreement rates between survey participants and LDA topics.
		}
		\centering
		\begin{tabular}{lrrr}
			\toprule
			\textbf{Answer type} &   \textbf{Train} &   \textbf{Test} &  \textbf{Test+rand} \\
			\midrule
			Top-1 only        &  65.1 &  59.2 &   68.0 \\
			Top-2 only        &  48.9 &  50.4 &   58.4 \\
			Top-3 only        &  36.3 &  39.3 &   8.2 \\
			Top-4 only        &  29.9 &  26.1 &   8.4 \\
			Top-1 or 2        &  83.2 &  81.9 &   88.4 \\
			Top-1 or 2 or 3   &  90.8 &  91.0 &  -- \\
			None of the above &   4.8 &   5.4 &   9.5 \\
			\bottomrule
		\end{tabular}
		\label{table:post_to_topic_results}
	\end{table}
	
	\subsection{Agreement rates for posts and topics} \label{ssec:topic_agreement}
	
	We report two metrics on the survey results: the post-level agreement rate and the topic-specific agreement rate.
	
	\header{Post-level agreement rate} is the percentage of answers for which the participant agrees with at least one of the designated topics of a certain type, aggregated over different participants. Here the types of choices are {\em Top-k only} (with $k=1, 2, 3, 4$), {\em Top 1 or 2}, {\em Top 1, 2 or 3}, or {\em None of the above}, presented as rows in \Cref{table:post_to_topic_results}. Agreements rates between the {\em train} and {\em test} settings are similar with a small decrease for answers in {\em test}, indicating that the topics generalize reasonably well to new posts. The decreasing trend from {\em top 1} to {\em top 4 only} is expected due to their decreasing LDA topic posteriors. In the {\em test+rand} setting, the presence of irrelevant (random) topics increases the probability that either the top-1 or top-2 topic being selected by 8\%, and {\em None of the above} by 4\%. This observation is consistent with well-known behavior patterns in choice-making \cite{simonson1992choice}, namely the tradeoff contrast that enhances options in the presence of unfavorable alternatives.
	
	\header{A topic-pair representation.} The average number of topics chosen by participants is 1.70 (\emph{train}: 1.80, \emph{test}: 1.75, \emph{test+rand}: 1.43). The frequencies for answer lengths can be found in Appendix Fig. E2. Given that the survey leaves the number of topics chosen unconstrained, this observation reveals that participants often perceive more than one topic being relevant to the post. Moreover, the agreement rate for {\em top 1 or 2 topics} is 81.9\% (+22.7\% from {\em top 1 only} and 9.1\% less than {\em top 1, 2 or 3}) for {\em test}, and {\em 88.4\%} on {\em test+rand}. This observation prompts us to define (unordered) {\bf\em topic pairs}, i.e., top-1 and top-2 topics for each post, as the automatically extracted relevant topics.  The topic pairs are unordered, because the posterior probabilities of top-1 and top-2 topics are close in value (\Cref{ssec:named_topics}). Additionally, as surfaced in the deliberation process of topic naming task (\Cref{ssec:naming}), annotators could not distinguish which of the top two topics is more prevalent. We posit that the topic-pair representation makes the classification of moral dilemmas significantly more nuanced and richer. Further observations on topic pairs are presented in \Cref{ssec:topic_pairs,section:observations}.
	
	\header{Topic-specific agreement rate}  is defined as the percentage of times that a given topic is selected when presented as either {\em top-$1$ or top-$2$} for a post, aggregated over different participants.  Results for {\em train} are shown in \cref{fig:topic_stats_train}, and those for {\em test} and {\em test+rand} are in Appendix Fig. E3, which show the same patterns for topic prevalence and agreement rate. We observe that frequent topics such as \topic{communication} and \topic{friends} have relatively higher agreement rates ($\geq 60\%$). Topics belonging to \emph{identities} generally have higher agreement rates than other meta-categories. A few infrequent topics have high agreement rates, such as \topic{pets}, which may be explained by being defined by animal related words. Topics such as \topic{jokes} and \topic{time} are among the least agreed upon; one explanation is that they may appear as the secondary topic or issue, together with another main issue. We note that the (weighted) average of the topic-specific agreement rates is lower than post-level agreement rate on the same setting, due to the latter requiring {\em either} the top-1 {\em or} top-2 topic being selected. 
	
	\begin{figure}
		\centering
		\includegraphics[width=0.8\linewidth]{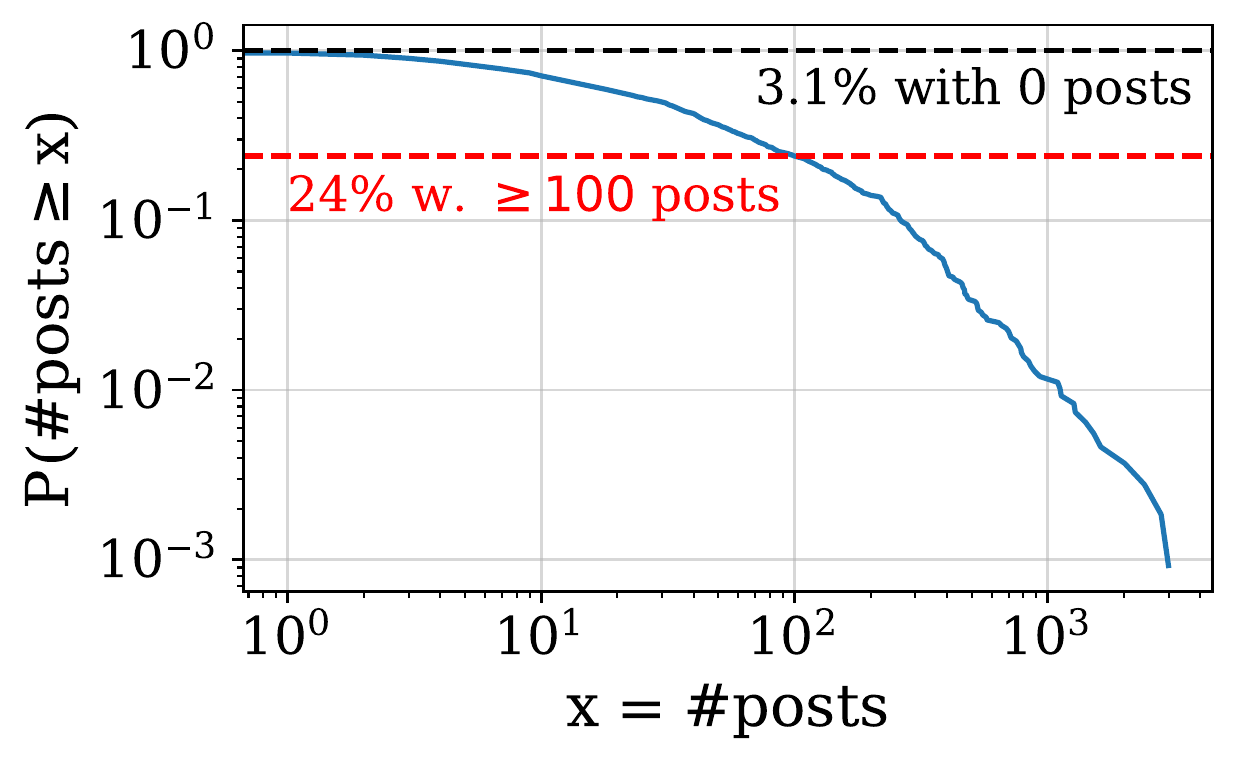}
		\caption{Distribution of topic pair sizes. The x-axis is the (log) size and the y-axis is the (log) CCDF.}
		\label{fig:pair_size_distribution}
	\end{figure}
	
	\subsection{A profile of topic pairs} \label{ssec:topic_pairs}
	
	From $47$ named topics, there are $\binom{47}{2} = 1,081$ unordered topic pairs. Among these, 33 pairs (3.1\%) have no posts, 396 pairs (36.6\%) contain at least 50 posts, and 259 pairs (24.0\%) contain at least 100 posts. The 10 largest topic pairs are shown in \cref{fig:topic_pair_empath} as row labels. \cref{fig:teaser} shows an overview of topic pairs, and \cref{fig:pair_size_distribution} shows the CCDF of size distribution for all topic pairs.
	
\begin{figure}[!t]
		\centering
		\includegraphics[width=1\linewidth]{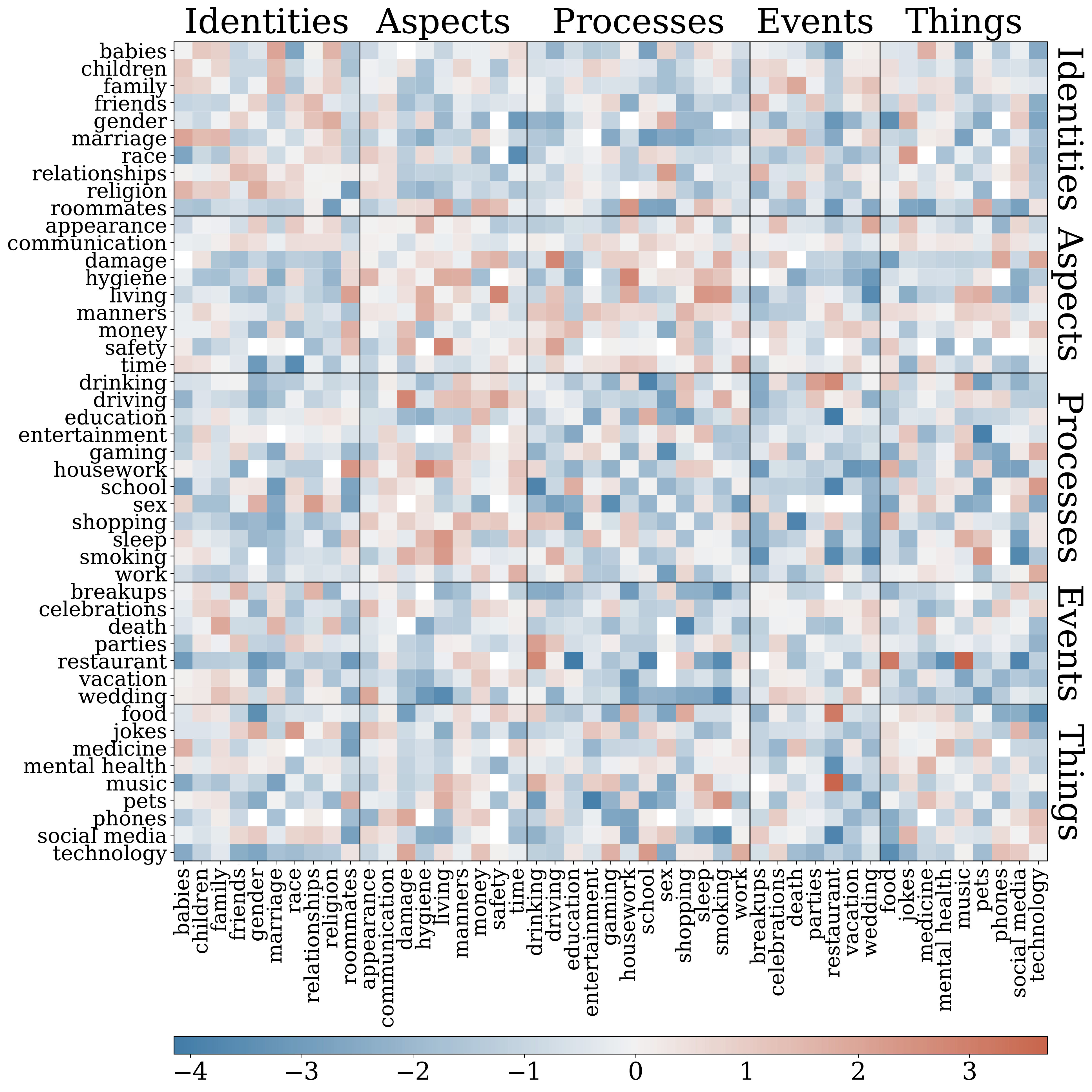}
		\caption{Point-wise mutual information between pairs of topics found by LDA, grouped by the five meta-categories.}
		\label{fig:pmi_lda}
	\end{figure}

	How often do we observe topics $k$ and $k'$ together? We use the point-wise mutual information (PMI) to quantify how much two topics co-occur more than prior-calibrated chance (PMI ~$>0$), or less than chance (PMI~$<0$):
	\begin{equation*}
		\mathrm{PMI}(k, k') = \log_2 \frac{p(k, k')}{p(k) p(k')}.
	\end{equation*}
	The PMI matrix is shown in \cref{fig:pmi_lda}. Among the meta-categories, topics in {\em identities} and {\em aspects} are likely to co-occur with another topic within the same meta-category, whereas those in {\em processes} do not. Topics in {\em aspects} tend to co-occur with those in {\em processes}, as one might expect. 
	
	Many topic pairs can be explained by semantic relatedness (or exclusion):  \topic{restaurant} tends to occur with \topic{food} and \topic{drinking} but not with \topic{education}.  On the other hand, some pairs appear to indicate conjunctions that are a frequent source of conflict and thus generate moral dilemmas. Some of these connections are obvious -- witness the high PMI for \topic{race} with \topic{jokes}, or \topic{children} with \topic{religion}. Both express domains that generate moral conflict on their own; one might reasonably expect even more conflict at their intersection. On the other hand, some conjunctions suggest more subtle patterns of conflict, like \topic{restaurant} with \topic{music}, or \topic{race} with \topic{food}. One or both of the topics in these pairs does not seem particularly morally laden on its own.  Some more complex interaction is likely at work. While the present work does not focus on particular mechanisms, we think this might be a rich topic for future work. We suspect that insofar as these pairs give rise to moral dilemmas, they might do so against a complex social background of expectations and norms. Additional profiles on the commenting and voting patterns across topic pairs can be found in Appendix Fig. E2.
	
	\section{Linguistic patterns in topic (pairs)} \label{section:observations}
	
	We examine the variations in word use across topics and topic pairs.
	
	\begin{figure*}
		\centering
		\includegraphics[width=1\linewidth]{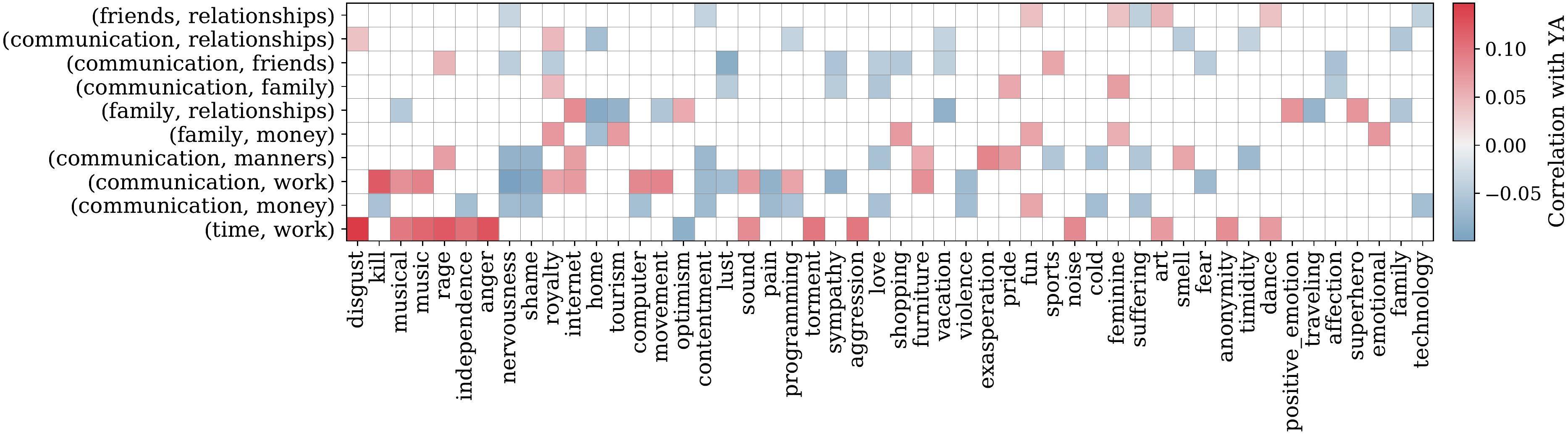}
		\caption{Pearson correlation between Empath categories and \YA in the 10 most frequent topic pairs. The columns are the top-50 Empath categories sorted by variance. White cells denote a lack of statistical significance ($p>0.05$, 388 in this plot).}
		\label{fig:topic_pair_empath}
	\end{figure*} 
	
	\header{Topic pair statistics via Empath categories} We first profile word use by Empath \cite{fastEmpathUnderstandingTopic2016}, a crowd-sourced collection of topical and subjective word lists, containing 194 categories (see column labels of \Cref{fig:topic_pair_empath} for examples) and 15 to 169 words in each category, totaling 7,643 words. We generate a 194-dimensional vector for each post, with elements corresponding to the fraction of words in each Empath category. For each topic pair, we compute the Pearson correlation between each Empath dimension and the binary indicators for \YA judgments. Results are presented in \cref{fig:topic_pair_empath}. Some categories, such as {\em independence}, negatively correlate with \YA in {\em (communication, money)} but positively correlate with \YA in {\em (time, work)}. Categories such as {\em love}, {\em shame}, {\em nervousness} negatively correlates with \YA in multiple topic pairs, whereas {\em fun} and {\em feminine} positively correlate with \YA in multiple topic pairs. These correlation patterns indicate that the moral valence of similar words may differ across different topic pairs. It also emphasizes that topic pairs are a key covariate for further analyses.
	
	\begin{figure*}[t]
		\centering
		\includegraphics[width=0.9\linewidth]{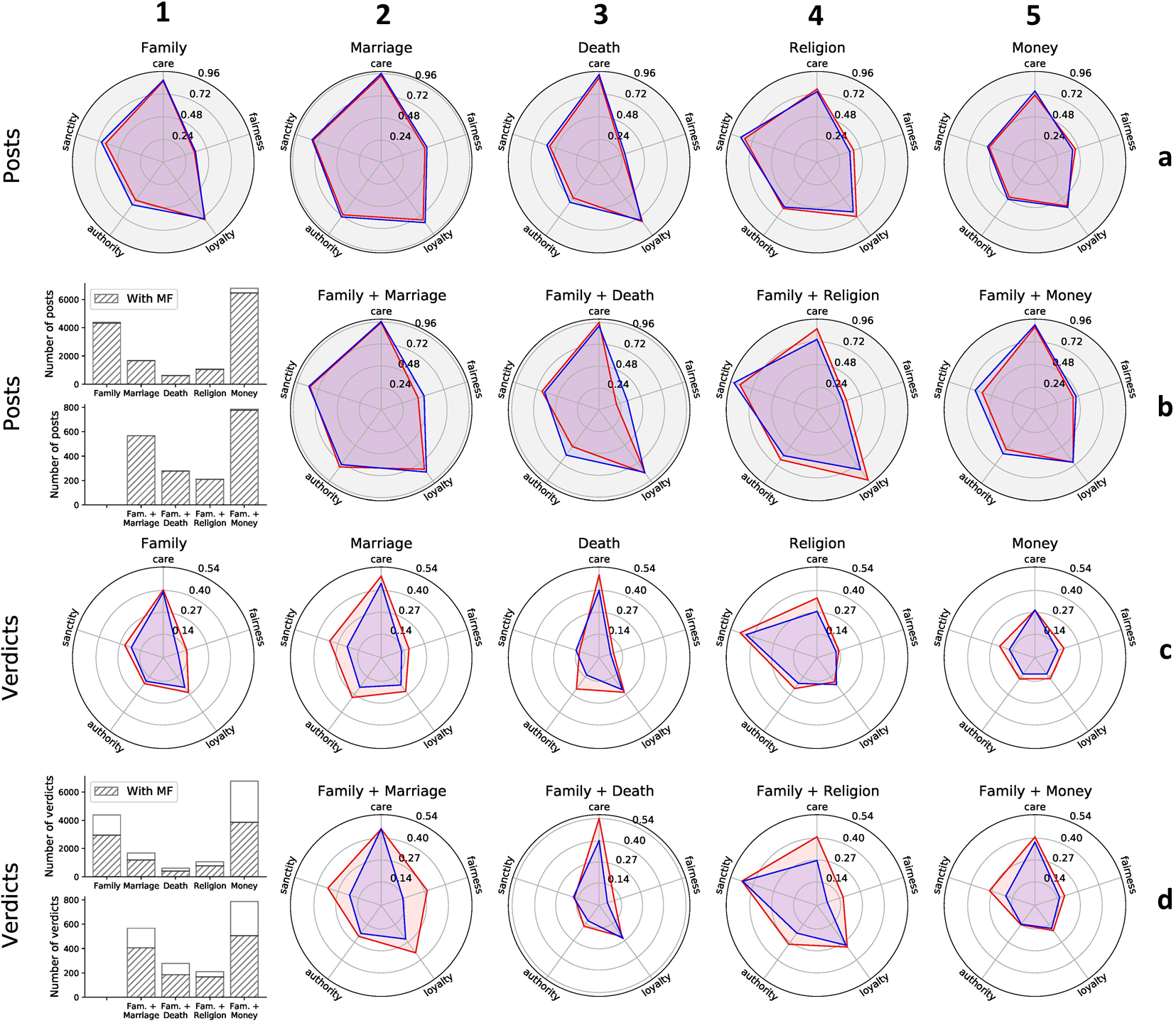}
		\caption{Strengths of moral foundations in selected topics and topic pairs. In each radar plot, each pentagon's vertex represents the proportion of posts (or verdicts) in a topic (pair) that have the presence of at least one moral foundation. Red (resp. blue) pentagons represent \YA-judged (resp. \NA-judged) posts (or verdicts). The bar plots at positions b-1 and d-1 represent the number of posts (or verdicts) in each topic (pair), and how many of them contain a moral foundation.}
		\label{fig:topic_mfd}
	\end{figure*} 
	
	\header{Scoring moral foundation axes}
	To directly examine the topics' moral content, we appeal to the widely used  Moral Foundations Dictionary (MFD), which projects the space of moral problems into five moral foundations: \textit{care}, \textit{fairness}, \textit{loyalty}, \textit{authority} and \textit{sanctity} \cite{haidtRighteousMindWhy2013}. We use the MFD 2.0 \cite{frimerMoralFoundationsDictionary2019a}, which contains 2,041 unique words in total. For each post, we compute a 5-dimensional binary vector, with each dimension being 1 if the post contains at least one word in the corresponding foundation. We do the same for the top-scoring comment of each post (called {\em verdict}). These vectors are aggregated over the posts/verdicts of the same topic or topic pair, and normalized by the total number of posts/verdicts. This yields a five dimensional vector with values between 0 and 1, representing the fraction of posts/verdicts with the corresponding foundation.
	
	\header{Moral foundation prevalence for topics and topic pairs} \cref{fig:topic_mfd} presents the proportions of posts (row a) and verdicts (row c) containing each foundation for five topics: \topic{family}, \topic{marriage}, \topic{death}, \topic{religion} and \topic{money}. We display these statistics for all topics in Appendix Figs. G1 and G2. Also in \cref{fig:topic_mfd} (rows b and d), the same proportions are presented for topic pairs involving \topic{family}. We observe some patterns consistent with the MFT. The foundation \textit{care} appears significantly in most posts of any topic: for example, nearly every post within the topic \topic{family} has the presence of \textit{care} (radar plot a-1). In posts about \topic{religion} (plot a-4), the authors tend to attach the foundations \textit{sanctity} and \textit{loyalty} in their narratives. These congruences provide a useful proof of concept. These observations are consistent when we look at the verdicts (plots c-1 and c-4, respectively). When topics are subdivided based on valence (\YA and \NA), the red and blue regions on row a mostly overlap, indicating there is little difference on what moral foundations positively and negatively judged posts appeal to. When we look at verdicts (row c), \YA verdicts typically adhere to every moral foundation more than \NA verdicts. This could be explained by the fact that negatively-judged comments are longer than positively-judged comments on average, increasing the chance they include a moral word.
	
	We also find evidence that secondary topics provide an interesting additional interpretive layer. \cref{fig:topic_mfd} (rows b and d) shows that the combination of topics often produces unexpected effects on the underlying moral foundations to which posts and verdicts appeal. For example, the combination of \topic{family} and \emph{money} produces \YA judgments that appeal to \textit{sanctity} more frequently than either topic does alone (plot d-5, compared with c-1 and c-5); a similar pattern is seen in \topic{family} and \topic{marriage} with \textit{fairness} (d-2, compared with c-1 and c-2). Conversely, some MFD loadings are driven more by one topic or another. The mechanism behind these interactions remains a topic for future research.  We suggest that this is good evidence that our topics provide a cross-cutting categorization \cite{DupreThe-disorder93} of the moral domain, one that might reveal more fine-grained structure that drives individual moral judgments. 
	
	There are also interesting dissociations between  posts and the verdicts. Posts use a wide range of identifiable moral language across different MFD domains. This confirms that posters to \AITA treat what they are saying as morally laden. The verdicts, on the other hand, tend to focus in on a smaller subset of moral considerations. For example, posters concerned with \topic{family} and \topic{religion} very often focus on both \textit{sanctity} and \textit{loyalty} (plot b-4), but verdicts tend to downplay that in favor of strong focus on \textit{sanctity} (plot d-4). Some reasons also seem to distinguish verdicts: \YA judgments for \topic{family} and \topic{marriage} focus more on \textit{loyalty} and \textit{sanctity} than do \NA judgments (plot d-2). These effects come apart from the original posts, where the radar plots largely overlap between \NA and \YA (plot b-2). These dissociations suggest that verdicts do not necessarily follow the original framing of the poster, and that the subsequent discussion plays an important role in focusing attention on details. They also show that dilemmas can have a non-additive structure \cite{kagan1988additive}, in which the presence of one topic can affect the importance of reasons raised by a different one. 
	
	Finally, we note that all five studied moral foundations are often present, to varying degrees, within what is broadly the same online population. Even strong predictable associations (such as \topic{religion} with \textit{sanctity}) coexist alongside appeals to other types of reasoning. It is no surprise that real-world cases are often quite messy. This is part of the attraction of AITA. Part of that complexity comes from the interaction of different domains, here revealed by our topics. Hence a bottom-up approach provides a valuable complement to experimental studies, which for good reason often focus on clear cases.  
	
	\header{Coverage of moral foundations dictionary}
	Of the 102,998 posts in the training data, there are only 5,425  (5.3\%) posts without the presence of any foundation in its description. However, we find that the MFD 2.0 has relatively low coverage on the \AITA verdicts. There are $44,282 ~(43\%)$ posts for which MFD finds no presence of any foundation in their verdicts. Of these, verdicts in topics \topic{phones}, \topic{music}, \topic{shopping}, \topic{roommates}, \topic{driving} and \topic{celebrations} have the highest missing rates of above $50\%$, while verdicts in \topic{religion} have the lowest rate of $28\%$. This is evidence that MFD 2.0 may miss important moral considerations, particularly on the comparatively shorter verdict posts. Below is an example verdict where the MFD 2.0 does not detect any foundation:
	
	{\sf\footnotesize
		\textbf{Post title}: ``AITA For Firing An Employee After His Parents Died?''
		
		\textbf{Verdict}: ``\yta for firing him without first going through the steps of describing his issues to him and giving him a chance to improve.  He's been back for only 2-3 weeks.  
		It's not about `having heart', it's about making a dumb business decision for both you and him.  So much smarter to work with this guy to get him back on track after a temporary setback than to push the eject button and have to find and start over with a new person. Dumb.''
	}
	
	This verdict appeals to considerations of both \textit{authority} and \textit{fairness}. \textit{Authority} is the power to issue commands and enact rules that are generally followed by the appropriate subject group; employer-employee relationships fall under this heading. \textit{Fairness} involves adhering to a set of procedural safeguards, and the employer in this example plausibly violated these procedures.
	
	We note that several versions of MFD have been introduced by different authors. Other types of lexicon are also available, such as the morality-as-cooperation vocabulary. Potentially combining different lexicons and validating our findings across different dictionaries are left as future work.
	
	\section{Conclusion} \label{section:conclusion}
	
	In this paper, we analyze more than 100,000 interpersonal moral dilemmas on a Reddit forum called \AITA. Using a multi-stage data-driven approach involving text clustering and human expert annotation, we group these posts into 47 high-quality topics with high coverage of 94\% of the dataset. Through crowd-sourced validation, we find high agreement between human annotators and our topic model when describing the themes of an \AITA post. Furthermore, we observe that topic pairs are better than individual topics at depicting a post's content, and therefore better serve as a thematic unit over \AITA posts. 
	We make several observations that suggest topic (pairs) is a key factor for thinking about daily moral situations. For instance, certain topics attract or repel other topics even when neither topic is particularly morally laden; the
	moral valence of similar words can vary across different topic pairs; and interaction effects in which final verdicts do not line up with the moral concerns in the original stories in any simple way.

	\subsubsection{Ethical considerations}
	We take steps to ensure that the study on moral dilemmas minimizes risk of harm.
	In both annotation tasks, we hide Reddit usernames and embedded URLs in posts to avoid identifying the original posters. We do not edit the names mentioned in posts since they are mostly initials or pseudonyms created by the poster. We present aggregated data that cannot be traced back to particular survey participants.
	Our survey design is approved by our institution's ethics committee. 
	
	\subsubsection{Limitations}
	
	As with all observational datasets, our collection method cannot retain posts and comments which had been deleted before the retrieval time, possibly leading to missing or incomplete data. Furthermore, it is impossible to precisely trace the comment containing the winning verdict in a thread, because after 18 hours (the amount of time after which the Reddit bot determines the flair), comments' scores can change drastically. This is a drawback compared to other Reddit datasets such as \texttt{r/ChangeMyView} in \citet{tanWinningArgumentsInteraction2016}, where the original posters explicitly give the winning comment a special symbol. Despite \AITA participants being self-selected, and cannot be considered a representative sample either of Reddit users or the population at large, this work assumes that the content in \AITA reflect daily life in interesting ways. The resulting topics provide evidence of the diversity and nuance of the set of daily moral discussion, and does not provide measures of representativeness for each topic. Our data is limited to posts that follow the posting guidelines set up by \AITA moderators. These guidelines prohibit posts about reproductive autonomy, revenge, violence, and conflicts with large social demographics. Conflicts within these prohibited topics could fall within the bounds of morality but are excluded from our dataset. 

	\subsubsection{Future directions}
	
	The present study only looks at posts and verdicts on \AITA. A natural extension would be to examine the content and structure of comments on each post. Our data also shows that posts often reflect a mixture of topics; it would be interesting to see whether the subsequent discussion preserves this mix or whether the search for reflective equilibrium \cite{RawlsA-Theory71} implies focusing on specific topics. It is also known that moral judgments can depend on the way situations are framed \cite{Sinnott-ArmstrongFraming08}; studying discussions might shed new light on these framing effects. An understanding of the extent to which everyday moral dilemmas on \AITA reflect the specific social or institutional roles embodied by its registered members could further demonstrate the usefulness of this domain on informing other moral decision-making tasks.
	
	Part of the motivation of studying \AITA was a philosophical interest in morally charged situations \cite{driverSuberogatory1992}. We are interested in the degree to which debates on \AITA might challenge the traditional distinction between moral norms and merely conventional norms like rules of etiquette \cite{FootMorality72,SouthwoodThe-moral/conventional11}. There have been recent challenges to this sharp division \cite{MartinI-think;90}. Our results are consistent with this challenge, with an important contribution from topics like \topic{manners} and \topic{communication} suggesting that the way things are done can be as important as what is done. Further work may shed light on what distinction, if any, can be drawn between the two domains. Second, a core tenet of early Confucian philosophy is that the everyday challenges and exchanges that people experience are of profound importance to morality \cite{OlberdingEtiquette:16}. We note that the everyday challenges and exchanges that occupied early Confucian philosophers are similar to our real-world moral dilemmas. Future research could help identify links between the two. Finally, we note that a large number of topics concern particular kinds of relationships, like \topic{children}, \topic{family}, and \topic{friends}. This may be of particular interest to care ethics, as well as some forms of virtue ethics and communitarianism, which emphasize the moral importance of meaningful relationships \citep{CollinsThe-core15}.

	\section*{Acknowledgments} \label{section:ack}
	This research is funded in part by the Australian Research Council Projects DP180101985 and DP190101507. Nicholas George Carroll is supported by the Australian Government Research Training Program (RTP) Scholarship. The authors would like to thank the Humanising Machine Intelligence Project, Alice Richardson and the anonymous reviewers for insightful comments and discussions, and the ARDC Nectar Research Cloud for providing computing resources.
	
	{\small
		\bibliography{AITA.bib}
	}
	
	\clearpage
	\appendix
	
	\counterwithin{figure}{section}
	\counterwithin{table}{section}
	\renewcommand\thefigure{\thesection\arabic{figure}}
	\renewcommand\thetable{\thesection\arabic{table}}
	
	\section{AITA: structure and winning verdicts}
	\label{appn:subreddit_structure}
	
	\begin{figure}[t]
		\centering
		\includegraphics[width=1\linewidth]{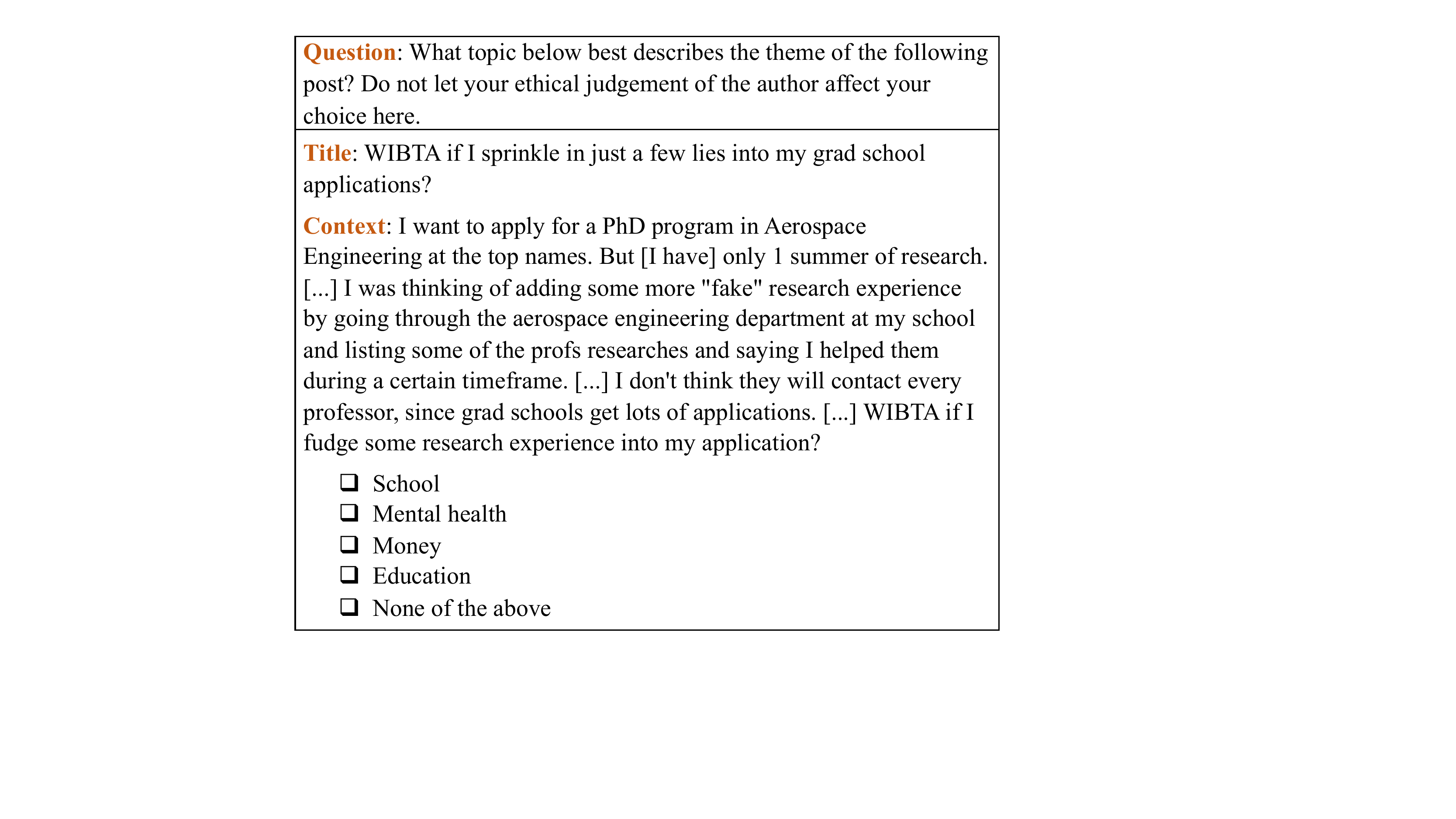}
		\caption{An example thread on \AITA containing a post (top) and comments (bottom). The gold badge indicates the most up-voted comment, which also becomes the winning verdict, \yta.}
		\label{fig:dataset:example_post_and_comments}
	\end{figure}
	
	The subreddit \aita (\AITA) is organized into \textit{threads}. \cref{fig:dataset:example_post_and_comments} gives an example thread in the dataset. On the top of the figure, the thread starts with a \textit{post} made by an \textit{original poster (OP)} or \textit{author}. The post contains a \textit{title}, its \textit{author's username}, \textit{posting time} and \textit{content} (or \textit{body text}). We only show the title and body text here. On AITA, titles should start with ``AITA'' (Am I the asshole) or ``WIBTA'' (Would I be the asshole), and the body text further describes the author's situation.
	
	Below each post are \textit{comments} made by the author or other people. Comments can also reply to other comments. A comment includes its \textit{author's username} and \textit{posting time}. In our dataset, a comment also has an ID, its parent's ID and the ID of the post it is replying to. To make a judgment, the comment's author must include one of the following five \textit{tags}: \yta (the OP is at fault), \nta (the OP is not at fault), \esh (everyone is at fault), \nah (no one is at fault) and \info (more information on the situation is needed to judge). In \cref{fig:dataset:example_post_and_comments}, we show two top-level comments (replying directly to the post) with different tags, and one lower-level comment (replying to another comment) without a tag.
	
	Posts and comments can be upvoted and downvoted by community members. The \textit{score} is the difference between upvotes and downvotes. To determine a post's verdict, after $18$ hours since the post, AITA uses a Reddit bot which tallies all comments below that post and uses the tag within the highest-scored comment as the winning judgment. That judgment is attached to the post as a \textit{flair}. In \cref{fig:dataset:example_post_and_comments}, the winning comment is shown with a gold badge, and the post is given an \yta flair.
	
	\begin{figure}[!t]
		\centering
		\includegraphics[width=0.6\linewidth]{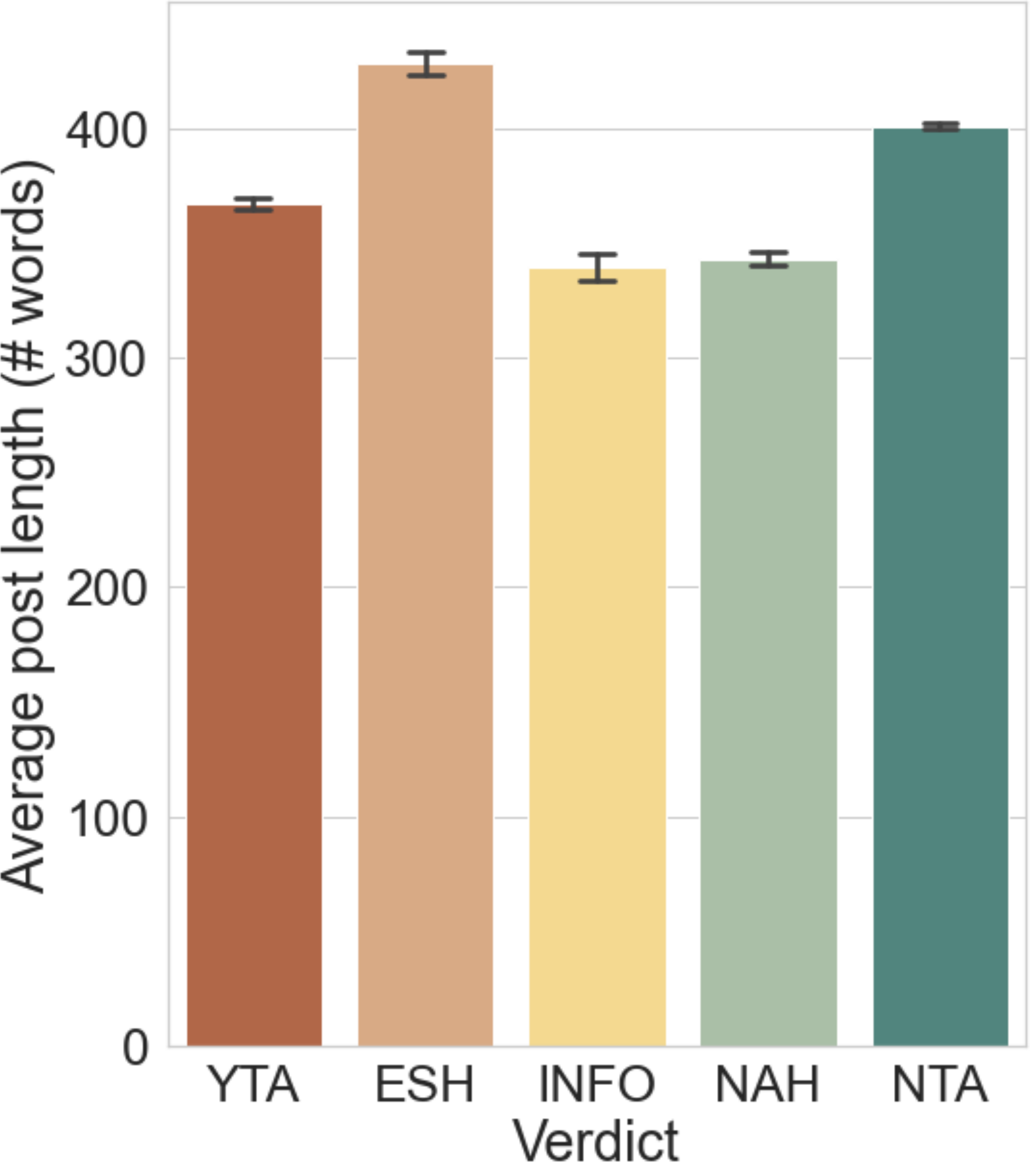}
		\caption{Average post length, grouped by flair} 
	\end{figure}
	
	\section{Initial topic exploration}
	\label{appn:topic_explore}
	
	In our first attempt to find categories in moral dilemmas, we performed a bottom-up discovery, inspired by card-sorting~\cite{spencer2009card}. The six authors of this paper each read the content of $20$ randomly selected posts (with each sample post read by at least $2$ people), and assign preliminary category labels (of one or a few words) according to their understanding. We are somewhat surprised to find that rather than falling into mutually exclusive categories as we initially anticipated (e.g., {\em money, relationship}), each post tends to have more than one label, which covers a wide range of aspects in daily life, such as {\em identities} (e.g., {\em friends, roommates}), {\em events} (e.g., {\em birthday, wedding}), and {\em themes} (e.g., {\em jealousy, dishonesty}). Moreover, on this small sample, the inter-annotator agreement is poor (around $30\%$), and the labeling practice among annotators varies from assigning a few broad labels to assigning a large number of detailed labels. Moreover, one author performs a {\em fill-in-missing-categories} exercise. It soon becomes clear that the number of categories that are not present in the small $60$-post sample but {\em could} be present can be vast, with the prevalence of it hard to estimate. For example, considering the topic of locations, the {\em school} label is present, but what about {\em gym}, {\em shops}, {\em restaurant}, {\em church}; and how many of each are there?
	
	This exploration leads us to conclude that manual discovery is not enough. In particular, the lack of definitions and a vocabulary for categories of moral dilemmas and assigning posts to categories prevents the progress on each other.
	
	\section{Topic modeling and text clustering}
	
	\subsection{Perplexity for LDA clusters}
	\label{appn:LDA_perplx}
	
	\begin{figure}[t]
		\centering
		\includegraphics[width=0.8\linewidth]{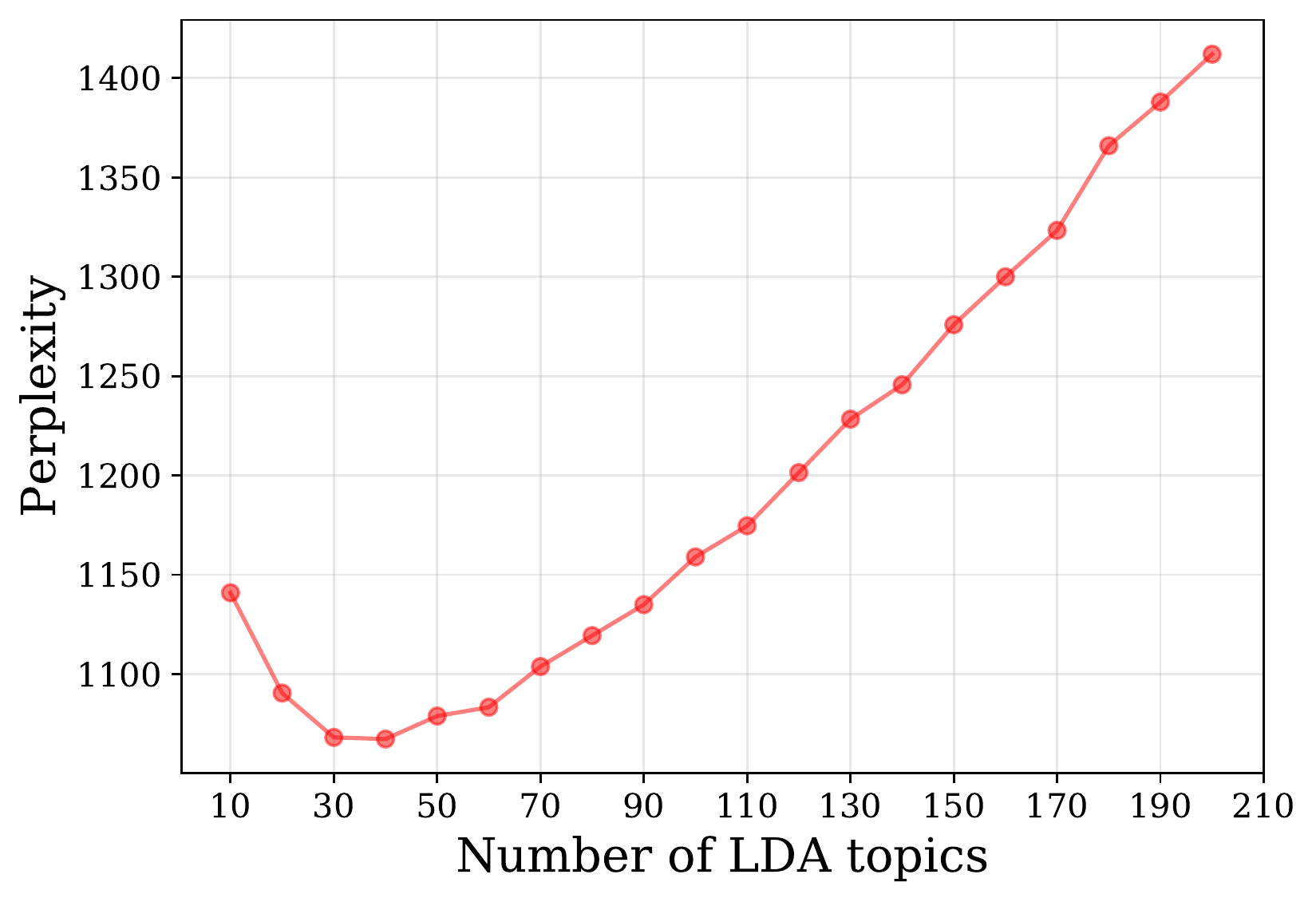}%
		\caption{Perplexity against the number of LDA topics.}
		\label{fig:topic_modeling:lda_perplexity}
	\end{figure}
	
	Choosing the number of clusters is a hyperparameter tuning task for LDA. To do so, we randomly split the $102,998$ posts into training and validation sets of ratio 80:20. We train LDA on the training set, and rely on the perplexity of the validation set to assess the number of topics. The perplexity is defined as
	\begin{align*}
		\text{Perplexity} = \exp\{-1 \times \text{log-likelihood per word}\},
	\end{align*}
	where the log-likelihood per word is estimated using its variational bound.
	
	We use the \texttt{scikit-learn} \cite{Pedregosa2011Sklearn} implementation of perplexity for the LDA model trained on the bag-of-words embedding. We refer to this model as \LW. \cref{fig:topic_modeling:lda_perplexity} plots the perplexity against a range of LDA topic numbers. The minimum perplexity is achieved at 40 topics.
	
	\subsection{Other text clustering methods}
	\label{appn:clustering_variants}
	
	Aside from \LW, we perform clustering using non-negative matrix factorization \cite[NMF]{paateroPositiveMatrixFactorization1994} and soft K-means clustering \cite{dunnFuzzyRelativeISODATA1973}.
	
	NMF performs a decomposition of a text embedding into a product two matrices, one for the document-topic distribution and one for the topic-word distribution. The former matrix is used to find the most salient topics for each document. We employ NMF on two embeddings: the $10,463$-dimensional TF-IDF and the $194$-dimensional Empath \cite{fastEmpathUnderstandingTopic2016}. The former has the same vocabulary as bag-of-words, but word counts are weighted by their inverse document frequency. The latter is an aggregate of words into a set of semantic categories. For example, the category \emph{social\_media} contains words like \emph{facebook} and \emph{twitter}. Informed by the LDA outputs, we set the number of clusters to $70$. We refer to the NMF model trained on TF-IDF as \NW, and that trained on Empath as \NE. 
	
	We also use soft K-means, with stiffness parameter $\beta = 1$, on the Sentence-RoBERTa \cite{reimersSentenceBERTSentenceEmbeddings2019} embedding to mimic topic modeling: each cluster is considered a topic, and the probabilities assigned to the topic for each document represent the likelihoods of the document belonging to the topics. Similarly, we also use $70$  clusters in training. We denote this model \KB.
	
	\subsection{Topic coherence}
	\label{appn:coherence_score}
	
	\begin{table}
		\centering
		\caption{Most and least coherent topics, along with their top word lists,
			for each model. The coherence score is the UMass metric.}
		\label{table:topic_classification:coherence}
		\begin{tabular}{lp{4.5cm}c}
			\toprule
			Model & Word list & \centercell{Coherence} \\
			\midrule
			\multicolumn{3}{@{} l}{\textit{~Most coherent clusters}} \\
			\centercell{\LW} & time, try, thing, life, help, year, like & -0.92 \\
			\centercell{\NW} & thing, time, like, love, try, year, month & -0.99 \\
			\centercell{\NE} & family, children, home, domestic\_work, wedding, party, friends & -0.22 \\
			\centercell{\KB} & class, time, school, work, like, day, friend & -0.74 \\
			\hline
			\multicolumn{3}{@{} l}{\textit{~Least coherent clusters}} \\
			\centercell{\LW} & account, ring, throwaway, propose, mate, bet, fiancée & -6.02 \\
			\centercell{\NW} & watch, movie, tv, video, film, netflix, laptop & -3.20 	 \\
			\centercell{\NE} & leader, law, government, power, royalty, dominant\_hierarchical, achievement & -1.83 \\
			\centercell{\KB} & come, work, try, number, time, interesting, obscure & -3.60 \\
			\bottomrule
		\end{tabular}
	\end{table}

	When displaying topics to users, each topic is generally represented as a list of $10$ keywords, in descending order of their topic-specific probabilities. Suppose that cluster $k$ is described by its word list $V_k$, of size $N$. For words $x_m$ and $x_l \in V_k$, let $D(x_m)$ be the number of documents in $k$ with at least one appearance of word $x_m$, and $D(x_m, x_l)$ be the number of documents containing one or more appearances of both $x_m$ and $x_l$. The topic coherence \cite{mimnoOptimizingSemanticCoherence2011}, often called the UMass coherence, of cluster $k$ is
	\begin{align*}
		C(k; V_k) = \sum_{m=2}^{N} \sum_{l=1}^{m-1} \log \frac{D(x_m, x_l) + 1}{D(x_m)}.
	\end{align*}
	The addition of the `smoothing' $1$ on the numerator is to avoid the logarithm of zero. Intuitively, the coherence metric quantifies the degree to which the keywords in a topic co-occur. More co-occurrences lead to a higher score (closer to zero), and correspond to a more coherent topic. An ‘ideal’ but clearly unrealistic case is the metric approaching zero (from below), with every document (in a very large collection) in the topic containing all the top words.
	
	We use an implementation of topic coherence called \texttt{tmtoolkit}.\footnote{\url{https://tmtoolkit.readthedocs.io}} For \LW, \NW, and \NE, the ordering of words within each topic is directly accessible. On the other hand, to mimic the importance-based ordering of the vocabulary for \KB, for each cluster $k$, we select the most frequent words. We use the $N=10$ words to calculate the coherence scores. Finally, to account for the effect of keyword list size, we divide each coherence score by $\frac{1}{2} N \times (N-1)$.
	
	\cref{table:topic_classification:coherence} presents the most and least coherent clusters, along with their 7 most salient words and topic coherence. Coherence scores for all \LW~ topics are in Table F1. We note that the least coherent cluster for \LW ~also is a cluster we named \topic{other} in Section 4.2 of the main paper.

	\subsection{Observations on clusters}
	
	\begin{figure}[t]
		\centering
		\includegraphics[width=0.8\linewidth]{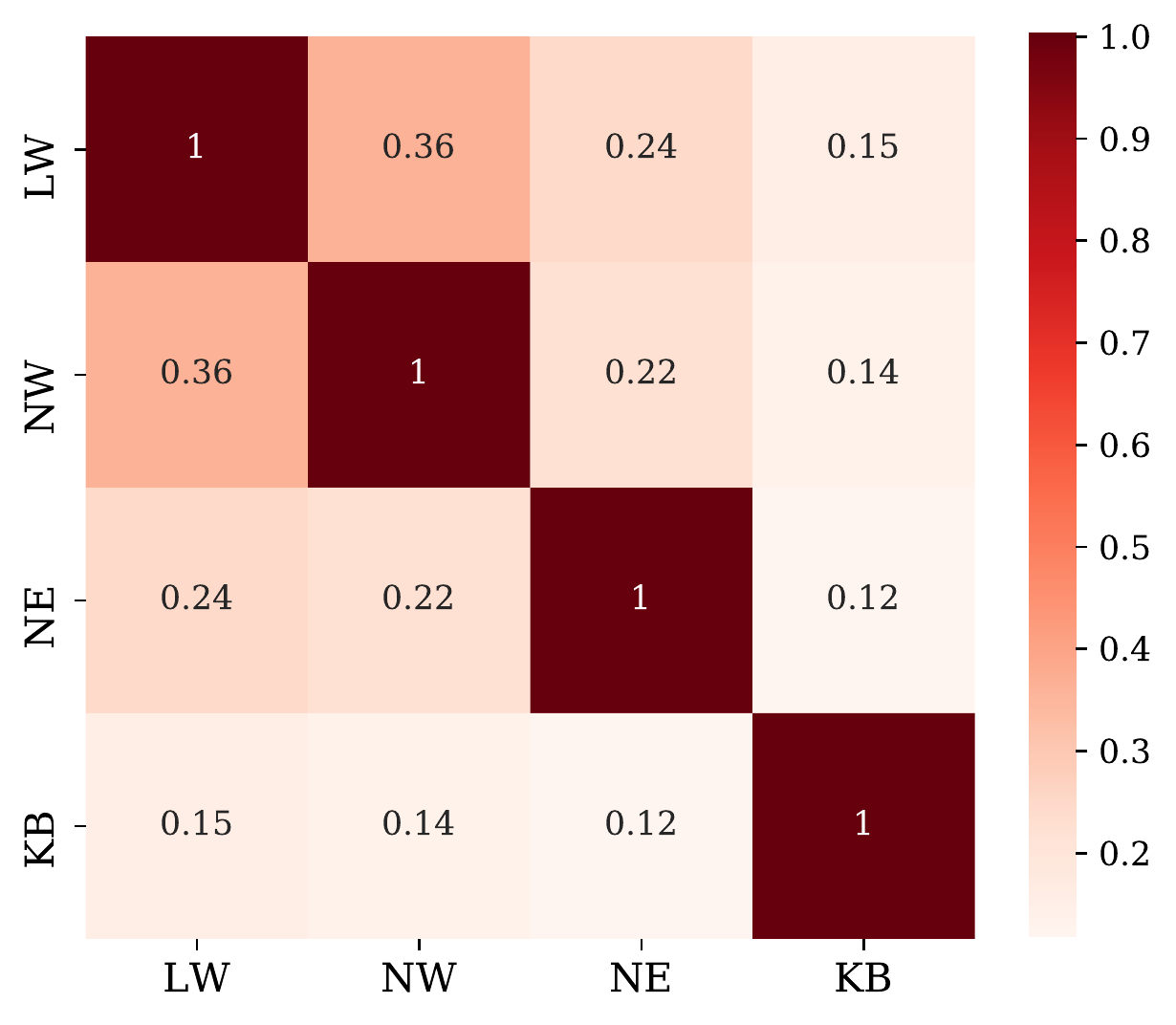}%
		\caption{Adjusted mutual information between clusterings.}
		\label{fig:topic_modeling:AMI}
	\end{figure}
	
	The \LW, \NW ~and \NE ~methods all create very uneven clusters, from those with mfewer than $50$ posts to those with nearly 10,000 posts. On the other hand, \KB ~clusters are mostly even in size, with most encompassing about $1.5\%$ of the posts.
	
	For \KB, while most clusters have similar sizes, the overlapping of clusters' content is common. Also, while each \LW ~cluster tends to match with one mother \NW ~cluster (in terms of their keyword lists), there are many \KB ~clusters that correspond to multiple
	\LW ~clusters. For example, the following \KB ~cluster with top-10 keywords \emph{(friend, like, date, talk, relationship, guy, year, time, feeling, girl)} tends to go with multiple other \LW ~and \NW ~clusters, including those about communication, relationships, and dating.
	
	We find that the Empath embedding used in \NE ~is useful in mdiscovering some mclusters, such as (\emph{work, business, occupation, white\_collar\_job, office, college, meeting, giving, school, blue\_collar\_job}). However, we find this embedding not informative enough, compared to bag-of-words and TF-IDF, in finding finer-grained topics. For example, an \NE ~cluster about communication co-occurs with \NW ~and \LW ~clusters on relationships, communication and social activities, making it too broad.
	
	Finally we observe that \LW ~tends to give the most coherent clusters on our dataset. \LW ~clusters are informative enough, in that their word lists are consistently related. They tend to correlate well with \NW ~clusters, but have more meaningful keywords.
	
	Considering each algorithm as a hard clustering by assigning each training example its cluster with the highest posterior probability, we compare the clusterings using the \emph{adjusted mutual information} (AMI) metric. Specifically, let $U$ and $V$ be two clusterings, the AMI between them is
	\begin{align*}
		\text{AMI}(U, V) = \frac{I(U, V) - \mathbf{E}[I(U, V)]}{\frac{1}{2}[H(U) + H(V)] - \mathbf{E}[I(U, V)]},
	\end{align*}
	where $H(U)$ is the entropy of $U$, $I(U, V)$ is the mutual information between $U$ and $V$, and $\mathbf{E}[I(U, V)]$ is the expected mutual information. AMI ranges from $0$ (no matching) to $1$ (perfect matching). See \cite{vinhInformationTheoreticMeasures2010} for more detail. \cref{fig:topic_modeling:AMI} presents the AMI between all pairs of clusterings. The largest AMI between two clusterings is that between \LW ~and \NW, corroborating our previous observation that \LW ~clusters tend to co-occur with \NW ~clusters. \KB ~clusters tend to match the least with other clusterings.
	
	\section{Survey for topic naming}
	\label{appn:naming_survey}
	
	In line with the literature on topic modeling
	\cite{boydgraberApplicationsTopicModels2017}, especially on LDA, we find that mnaming a cluster should not simply be based on the cluster's most salient keywords. As pointed out in the following survey, we find some clusters having somewhat related words to humans but, after carefully reading some of their representative posts, we decided that they should not be considered topics. As a result, we set up some criteria for a cluster to be given a topic name.
	
	To be considered a topic, a cluster should have the following properties:
	\begin{itemize}
		\item its keyword list should be unambiguous in suggesting the topic's theme;
		\item its posts should be about the same topic when read by humans; and
		\item ideally, several human readers should agree on the topic name, given its keyword list and some posts.
	\end{itemize}
	To this end, we organized an annotation task among the six authors of this paper, asking ourselves to give a short (one- to two-word long) name for each cluster. As described in Section 4.2 of the main paper, we used the 70 clusters found by LDA, resulting in $70$ questions in total. Below we describe the components of this survey in more detail.
	
	\subsection{Choosing the keywords for each cluster}
	In this topic model, each cluster is described by a probability distribution $p(x \mid k)$ over the entire vocabulary. More salient words are given higher probabilities. The vocabulary is sorted by this probability to find the top words for each cluster. We choose the top $10$ words to describe a cluster, which is in line with other work in the literature \cite{boydgraberApplicationsTopicModels2017} (and references herein). In addition, we find that on average, only 11 of each cluster's top words have $p(x \mid k) > 10^{-2}$. \cref{fig:top1vstop2} shows the top-$1$ and top-$2$ cluster probabilities for some topics. \cref{table:LDA_clusters_full} lists all $70$ clusters along with their top-10 keywords. The keywords on each row are presented in decreasing order of topic-word probability.
	
	\subsection{Choosing the example posts for each cluster}
	As described in Section 4.2, the cluster sizes range from $16$ to $7,855$ posts, over the entire $102,998$ posts in the training set. A very small number of posts might not sufficiently describe a cluster, while a very large number is not feasible for a human annotation task. We decide to use six posts to describe each cluster. Three of the posts are chosen randomly from the posts with the highest topic-post posterior probabilities; we call them \emph{clear} posts. The other three are chosen randomly from posts with the lowest posteriors, and are called \emph{mixed} posts. This task was done to ensure that the post lists contain both posts which may seem clear to annotators about their topic name, and posts which may confuse the annotators.

	\begin{figure}[t]
		\centering
		\includegraphics[width=0.98\linewidth]{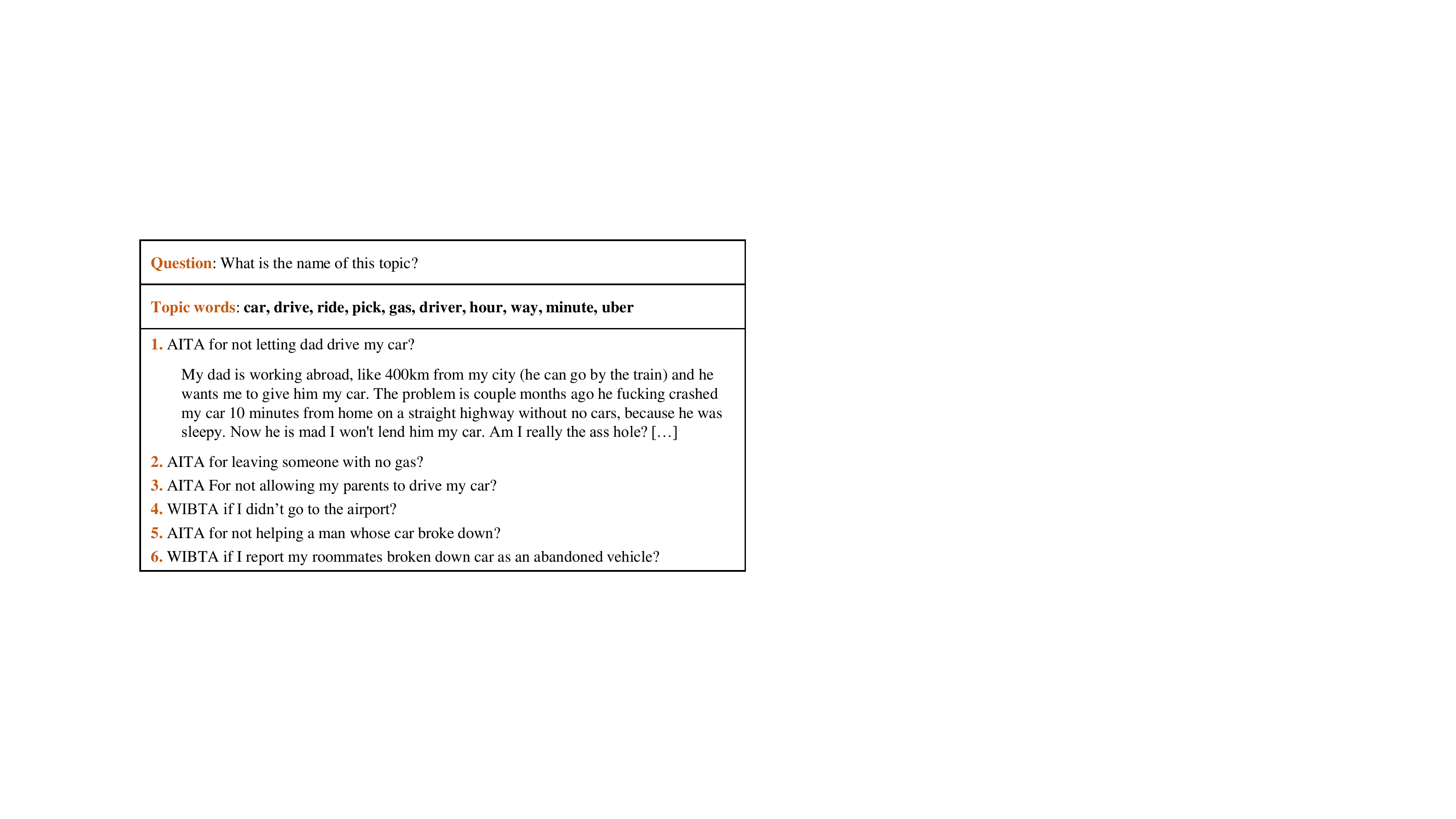}
		\caption{Example question in the cluster naming survey. The annotator is
			given a list of topic words (in bold) and six posts. Each post consists of
			its title (in blue) and body text which can be revealed by clicking on the
			title. Due to space constraint, only one post is shown here. The annotator
			is asked to provide a name to this topic in one or two words.}
		\label{fig:surveys:topicnames}
	\end{figure}
	
	\subsection{Question format}
	An example question is given in \cref{fig:surveys:topicnames}. First, we present the cluster's top-10 keywords as a list on top. Then each of the cluster's six posts is given below it. In each post, we present the post's title first, then followed by its body text. We only show the first 100 words of the post's body to ensure the six posts do not take over too much space in each question. We made one observation that the first 100 words are sufficient in describing the post's content. If the text exceeds this threshold, we simply use ``[\ldots]'' to denote the rest. To save space, we only show the body text of the first post in \cref{fig:surveys:topicnames}. Finally, The text box where the annotators give their answers at the bottom of the question is omitted in the figure.
	
	\subsection{Organizing annotation among authors}
	Six authors of this paper participate in this annotation task. We design the survey using Qualtrics.\footnote{\url{https://www.qualtrics.com/}} To avoid long survey time, we allow each author to participate multiple times, each with $10$ randomly chosen questions. If an author gives more than one answer to the same question (due to randomly chosen questions), only the earliest answer is kept. We keep participating until $3$ answers from $3$ different authors are recorded for each question, resulting in $210$ answers in total.

	\begin{figure}[t]
		\centering
		\subfloat[]{%
			\includegraphics[width=0.5\linewidth]{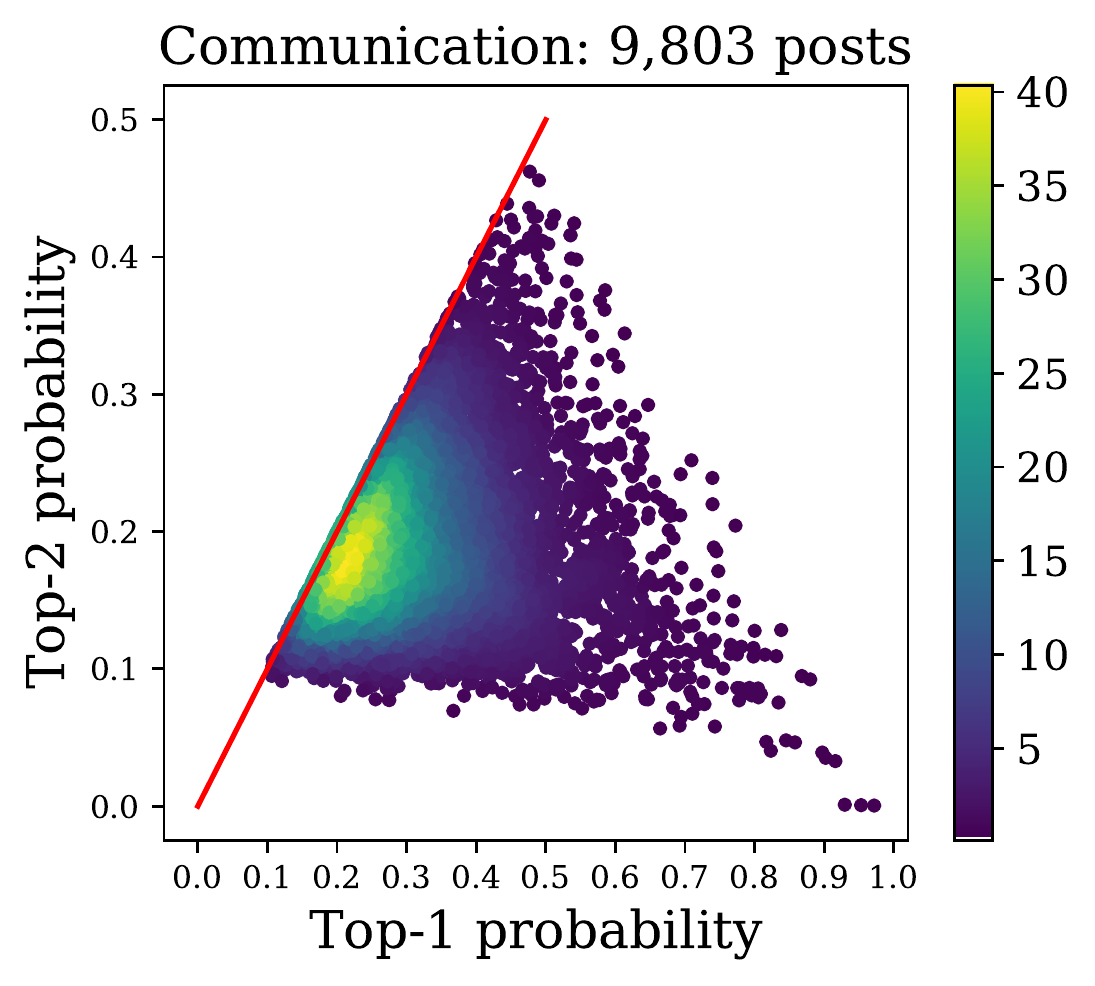}%
		}
		\subfloat[]{%
			\includegraphics[width=0.5\linewidth]{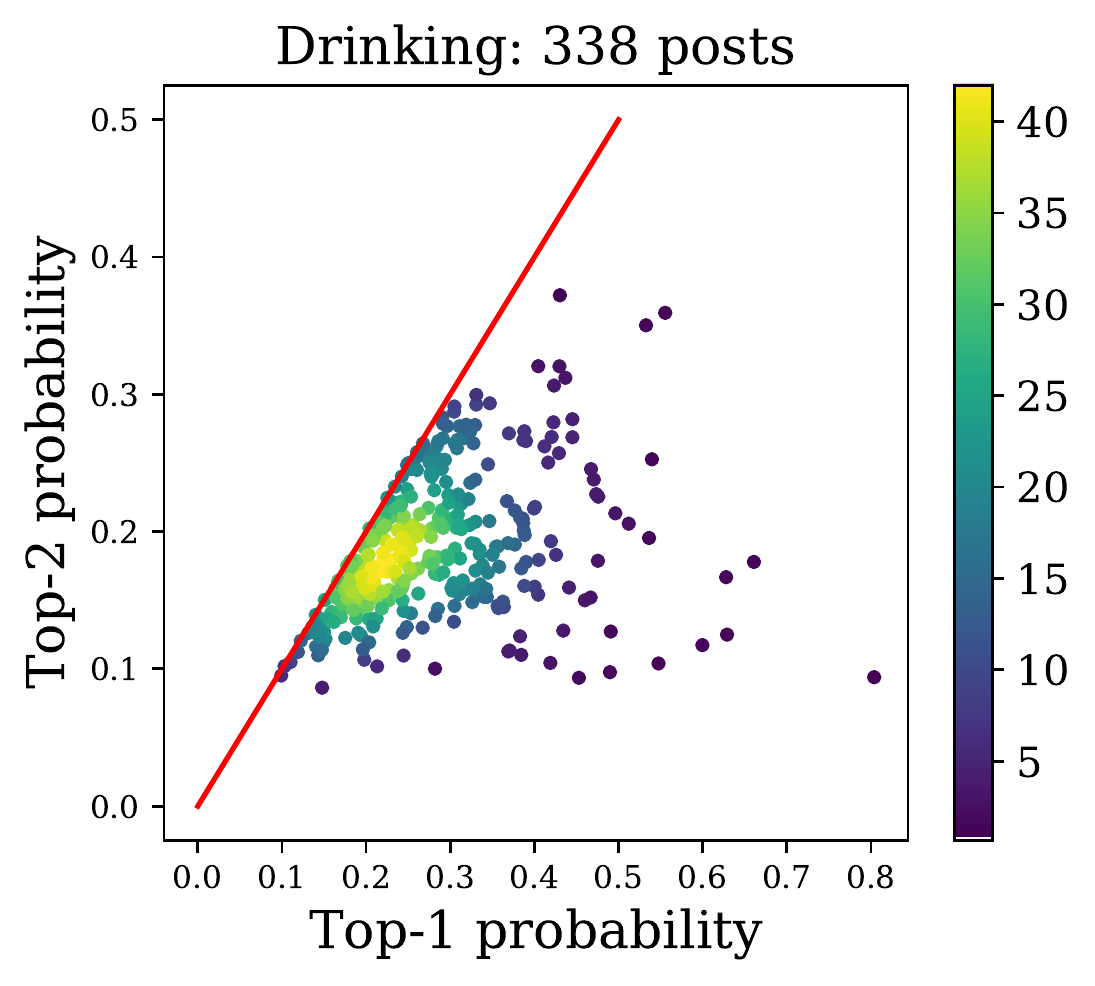}%
		}
		\\
		\subfloat[]{%
			\includegraphics[width=0.5\linewidth]{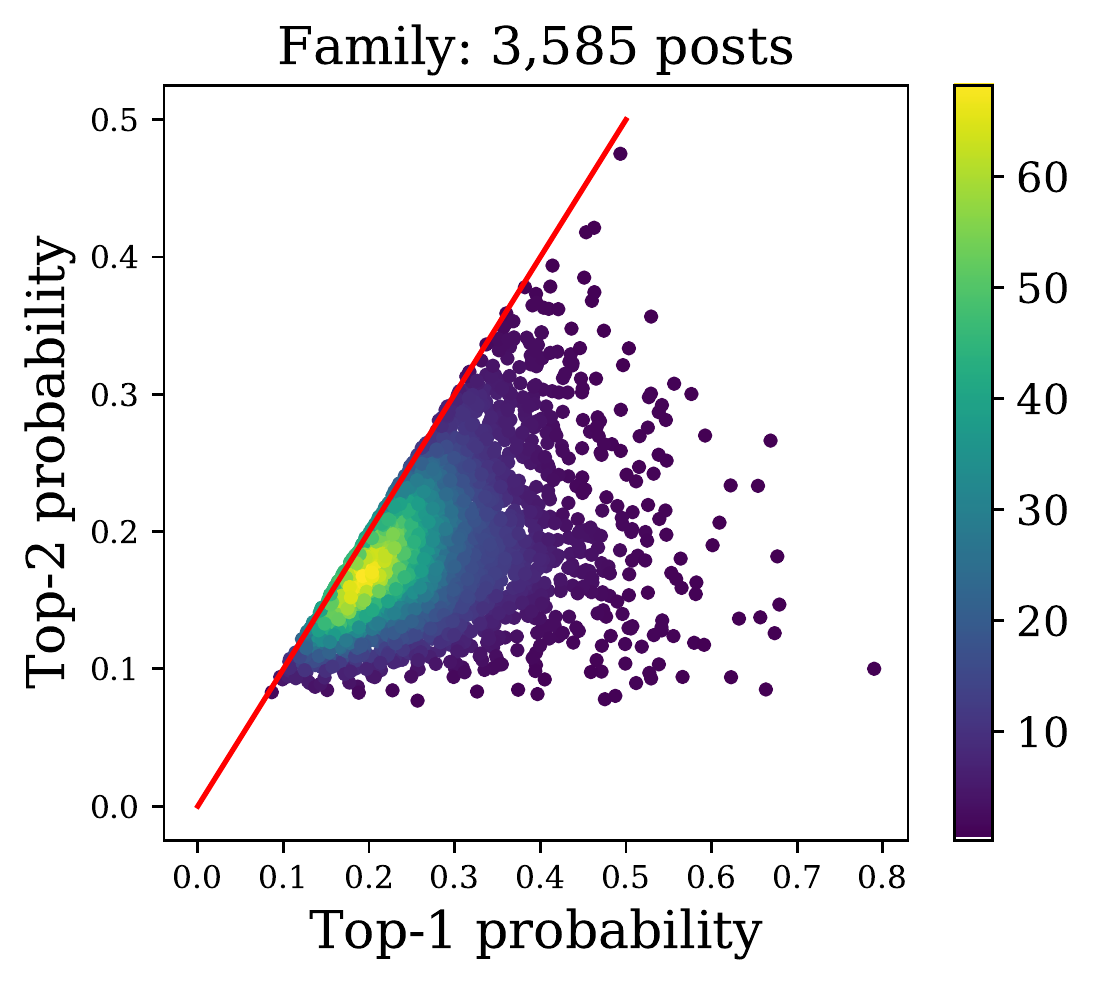}
		}
		\subfloat[]{%
			\includegraphics[width=0.5\linewidth]{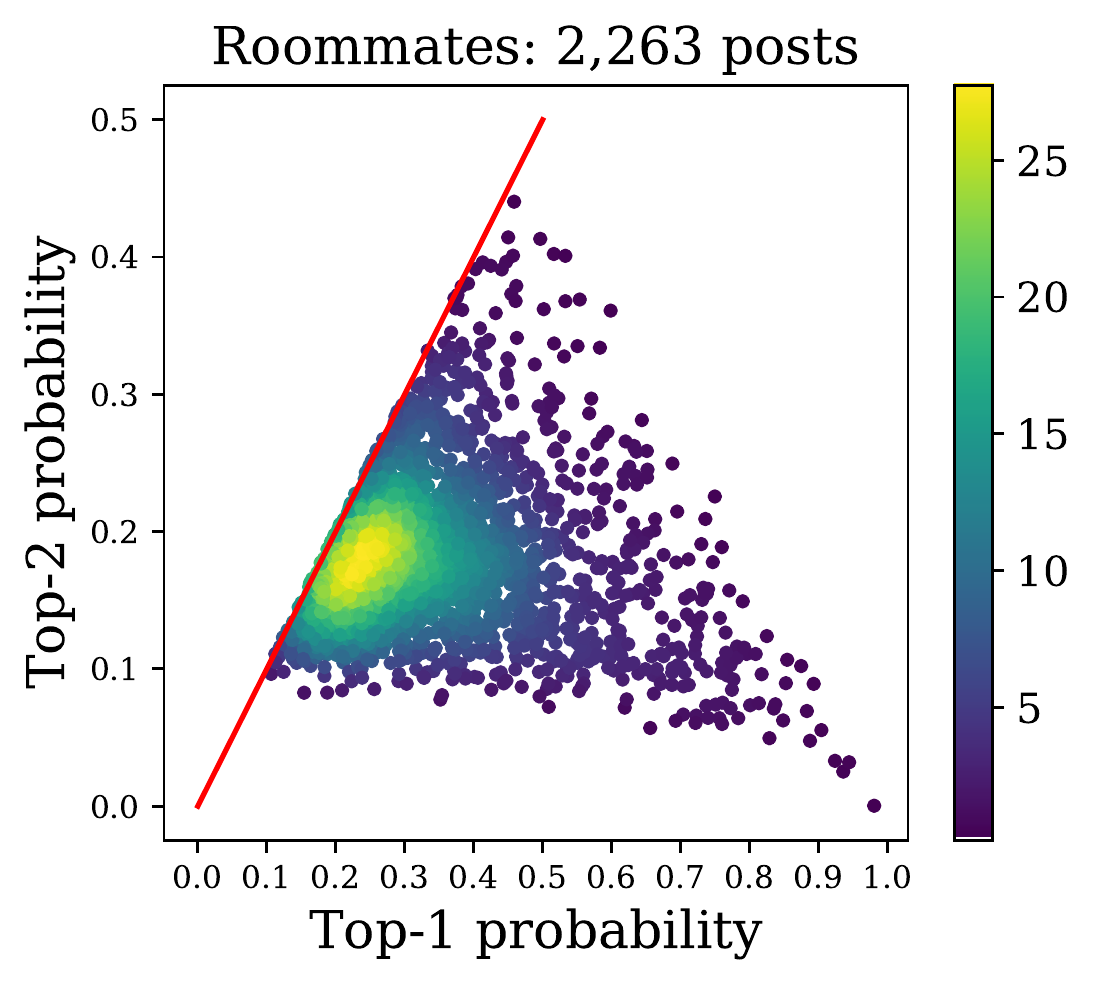}%
		}
		\caption{Scatter plot with kernel density estimation of the top-$1$ and top-$2$ LDA probabilities of posts in some topics.} 
	\label{fig:top1vstop2}
\end{figure}

\subsection{Post-annotation discussion of results} We determine that consensus is reached if at least $2$ of the $3$ answers for each question agree. Two authors met to discuss the survey results and categorized the consensus of the 3 answers in each question into four types:
\begin{itemize}
	\item \textit{Unanimous}: all $3$ answers are identical. The topic name is set to the answer. There are $17$ clusters of this type.
	
	\item \textit{Wording}: all $3$ answers are synonymous or are very close in meaning. An example is (\topic{celebration}, \topic{gifts}, \topic{celebrations}), after which the topic is named \topic{celebrations}. There are $41$ clusters of this type.
	
	\item \textit{Deliberation}: there is agreement between at least $2$ answers, but after carefully looking at the topic, we decided to rename it. An example is (\topic{family}, \topic{family}, \topic{family}), which initially was named \topic{family} but later changed to \topic{death}, because most of the posts in this cluster are about the passing of family members. There are $9$ clusters of this type.
	
	\item \textit{Other}: there is disagreement among the $3$ topics. For instance, one question received (\topic{entertain}, \topic{relationships}, \topic{army}). We decided to name all these clusters \topic{other}. There are $3$ clusters of this type.
\end{itemize}

\subsection{Results and discussions} After revising the clusters' names, we end up with $47$ named topics and one placeholder topic called \topic{other}. The total number of posts with a named topic is $96,263$, accounting for $93.5\%$ of posts in the training set. The topics were reduced from $70$ to $48$ because of overlapping names. We merge clusters with the same name together into a \emph{topic}. The highest number of overlapping names is $5$, for topic \topic{family}. \cref{table:LDA_clusters_full} lists all clusters, their names and sizes as a number of posts in the training set. \cref{table:LDA_clusters_full} lists all LDA clusters, described by their keyword lists, and their sizes along with the names found above.

We note that because of a considerable number of clusters falling into the \textit{wording} and \textit{deliberate} categories, it is easy that one annotator alone comes up with a different set of topic names, or considers a cluster as a meaningful topic when it should not be. The post-annotation step, conducted by more than one annotator, is an important part of this task.

\section{Crowd-sourced validation of topics}
\label{appn:post_to_topic}

To assess how well the topics found in Section 4.2 describe a post's content, we design a crowd-sourced study with multiple settings to verify the names for a large number of posts in the dataset. This section provides more additional information to that described in Sections 5.2 and 5.3.

\subsection{Question format}
Each question provides participants with one post, and five answers for the topic of the post. An example question in the survey is found in Fig. 5 in the main paper. Each question contains a prompt: \textit{``What topics below best describe the theme of the following post? Do not let your ethical judgement of the author affect your choices here.''} Below the prompt is the question, starting with its title and body text (which is called ``context'' in the figure). For brevity, we omit some details of the example post's body text. Finally, the five options appear at the end of the survey. The first four options are topic names found in Section 4.2. A participant can choose one or more topics in the first four options. The final option is \textit{None of the above}, which the participant can choose when the no topic satisfactorily describes the given post's theme.

\subsection{Choosing the posts for the questions}
We focus on the 47 named topics in Section 4.2, omitting the topic \topic{other}. To ensure that each topic has posts in the survey, we randomly sample posts in each topic. We initially conduct two surveys of the same format but with posts from the training set and test set:
\begin{itemize}
	\item Training set: These are posts in the training set (posts before 2020) for LDA in Section 4, of size 102,998. For each topic $k$ of the 47 topics, we randomly select 20 posts whose most probable topic is $k$ (based on posterior probability). The result is $47 \times 20 = 940$ posts, or 940 questions in the question bank. We call this survey setting \textbf{train} in Section 5.1.
	\item Test set: These are posts in the test set (posts in the first four months of 2020), of size $5,294$. For each topic $k$, we randomly select 10 posts. There are $5$ topics with fewer than 10 posts, resulting in only $450$ posts in the question bank, as opposed to $47 \times 10 = 470$ posts. We call this survey \textbf{test} in Section 5.1.
\end{itemize}
Doing so guarantees that every topic has posts in the survey, and allows us to compare the agreement rates among topics later on.

\subsection{Choosing answers for each question}
For each post in the \textbf{train} and \textbf{test} surveys, we choose the four most probable topics, according to the LDA posterior, for that post. If any of these four topics happens to be the topic \topic{other}, we remove it and add the fifth most probable topic to the list.

In Section 5, we also examine the effect of replacing the top-3 and top-4 topics with randomly chosen topics. To do so, we use the 450 posts in the \textbf{test} survey, and for each post, the top 2 LDA topics are kept, while the other two are randomly chosen from the rest. This creates a new survey setting, which we call \textbf{test+rand} in Section 5.1.

Finally, in each question, the four topics are randomly ordered in the answers.

\subsection{Survey setup and implementation.}
\label{appn:survey_implm}

\begin{figure}[t]
	\centering
	\subfloat[]{%
		\includegraphics[height=0.47\linewidth]{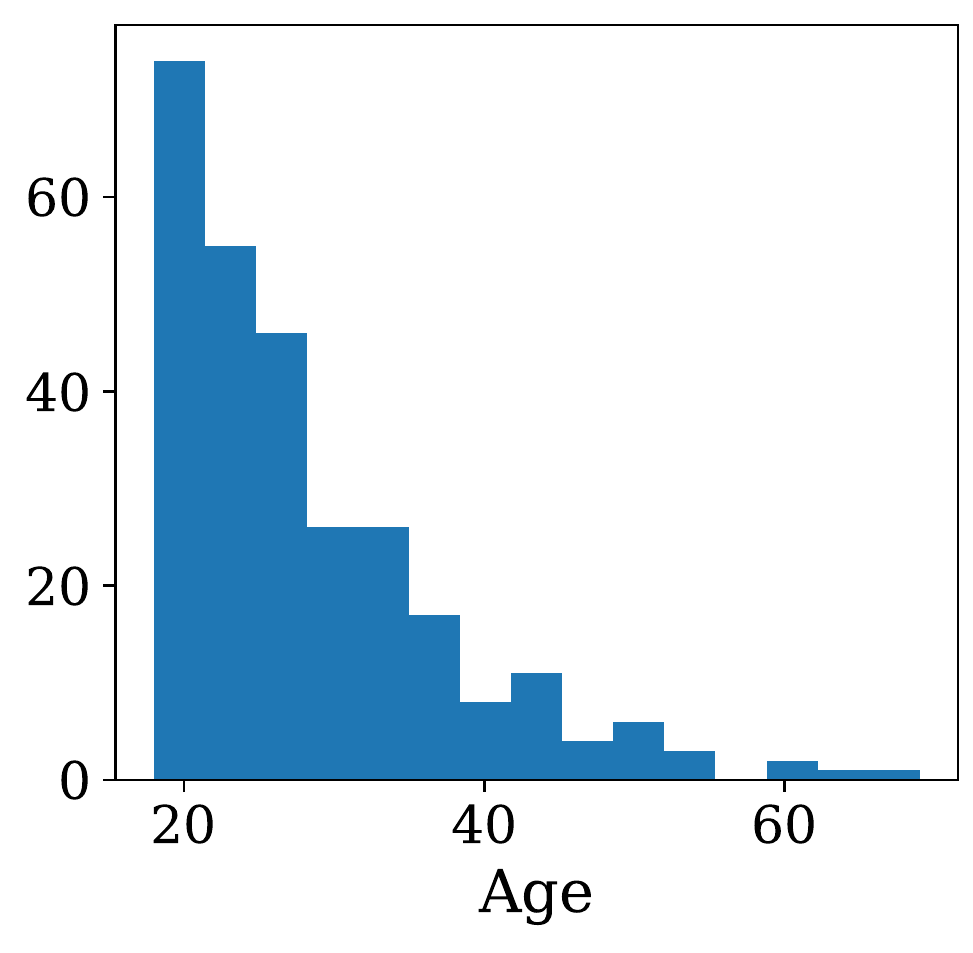}%
		\label{fig:crowdsourcing:demo:age}
	}
	\
	\subfloat[]{%
		\includegraphics[height=0.47\linewidth]{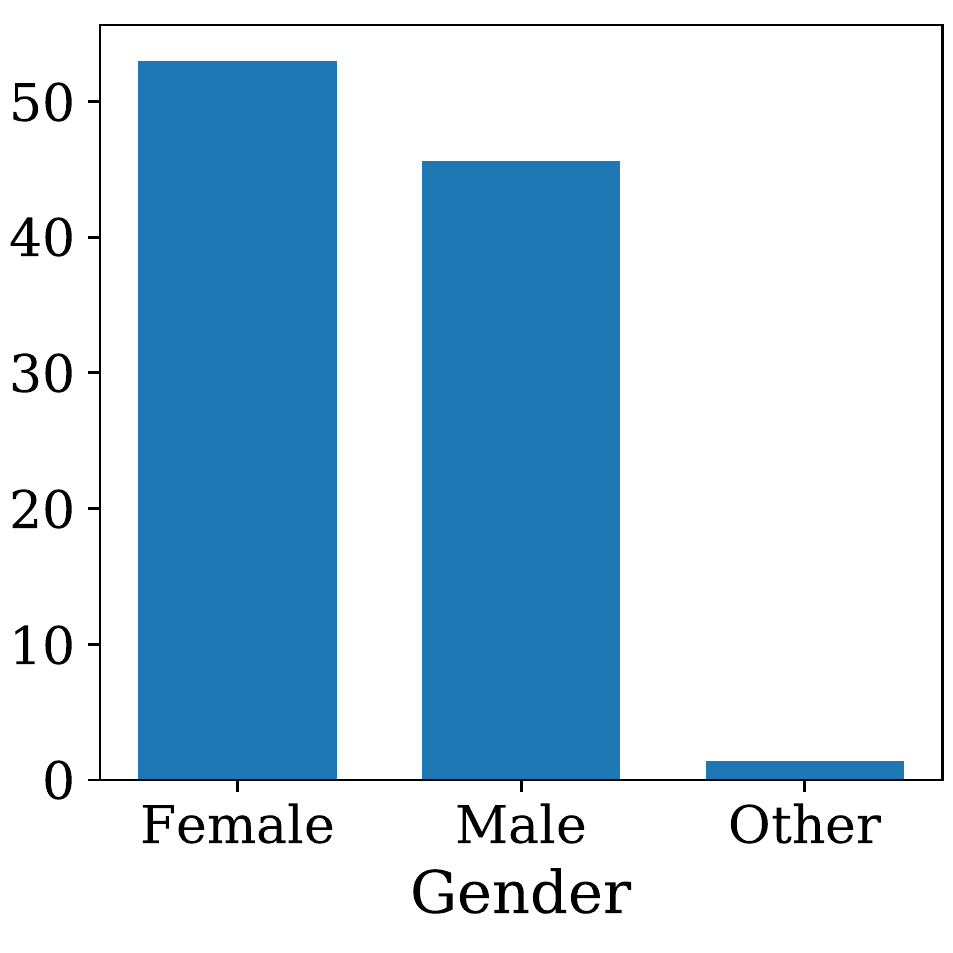}%
		\label{fig:crowdsourcing:demo:gender}
	}
	\
	\subfloat[]{%
		\includegraphics[height=0.47\linewidth]{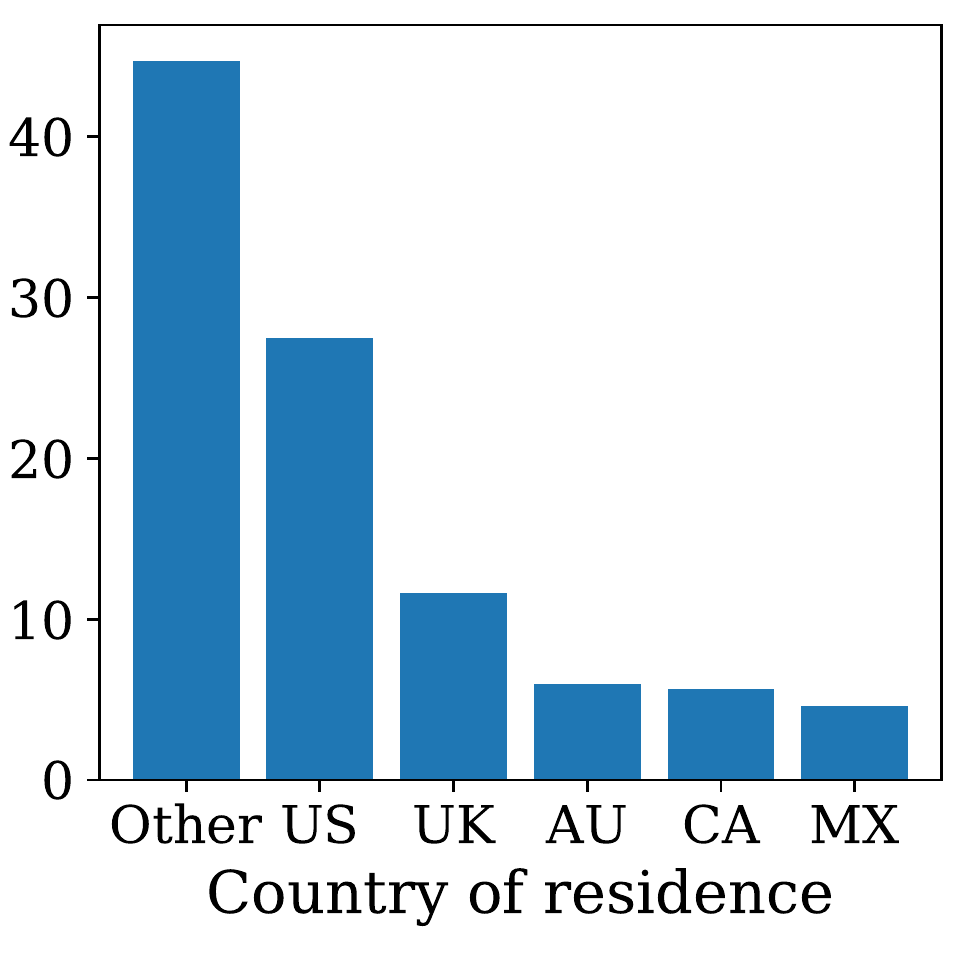}%
		\label{fig:crowdsourcing:demo:residence}
	}
	\subfloat[]{%
		\includegraphics[height=0.47\linewidth]{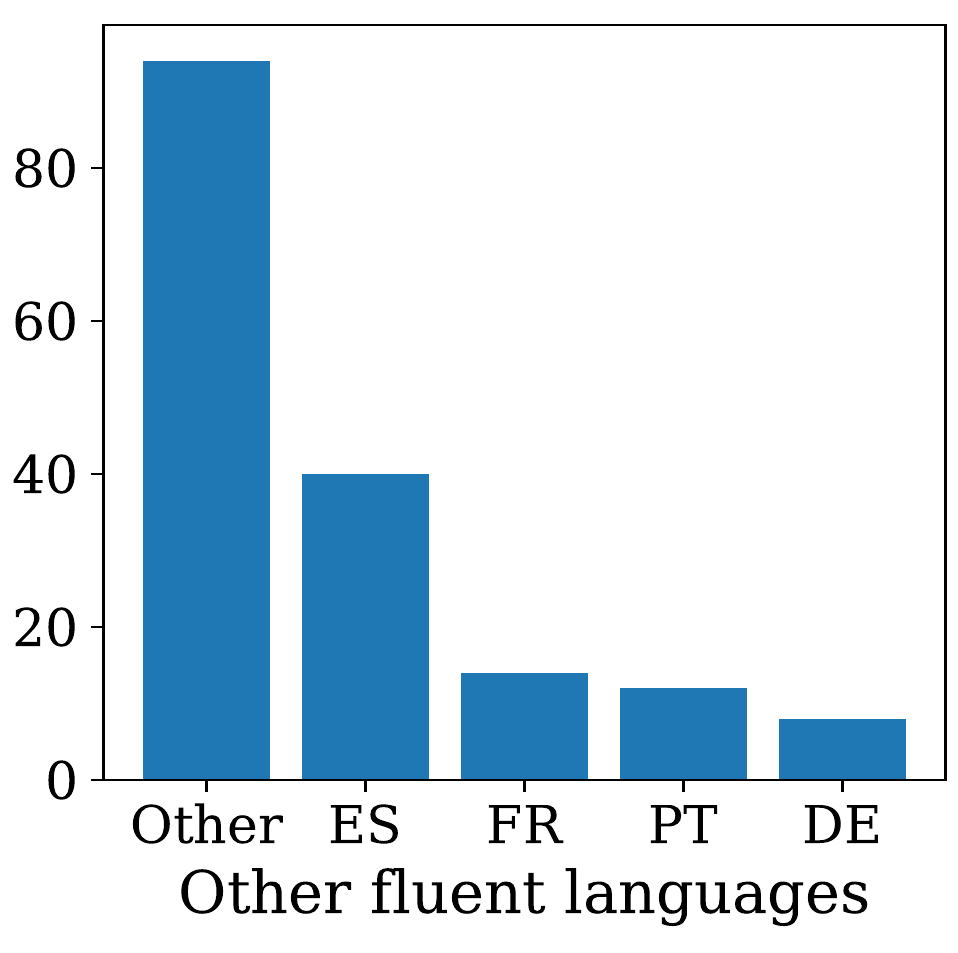}%
		\label{fig:crowdsourcing:demo:langs}
	}
	\caption{Demographic information of the 285 participants on Prolific. \protect\subref{fig:crowdsourcing:demo:age} Distribution of participants' ages. \protect\subref{fig:crowdsourcing:demo:gender} Gender shares. \protect\subref{fig:crowdsourcing:demo:residence} Most popular countries of residence. \protect\subref{fig:crowdsourcing:demo:langs} Languages they are fluent in, other than English.}
	\label{fig:crowdsourcing:demographic}
\end{figure}

We custom-design the survey using the SurveyJS library\footnote{\url{https://surveyjs.io}} and host the website on Heroku.\footnote{\url{https://heroku.com}} To recruit participants, we use the Prolific platform\footnote{\url{https://www.prolific.co}}. We record participants' demographic information, including their age, gender, residence, and fluent languages. We enforce one entry requirement that each participant must be fluent in English. A summary of this information is given in \cref{fig:crowdsourcing:demographic}.

Thanks to a large number of participants on Profilic, we could allow each participant to enter the survey once. Each participant is given 20 randomly chosen questions from the question bank. As a reminder, there are three question banks, \textbf{train}, \textbf{test} and \textbf{test+rand}, as described previously. We recruit participants until all questions are answered 3 times by 3 different individuals. The remuneration is \textsterling 2.5 for each participant, which averages to \textsterling 0.125 per question. We expect participants to take 20 minutes each time, resulting in the average pay of \textsterling 7.5/hour. This amount is recommended by Prolific, and is higher than the minimum pay of \textsterling 5/hour. Finally, this survey received ethical approval from the authors' institution.

\cref{fig:crowdsourcing:demographic} shows some demographic information about participants. In total, we recruit 285 participants, with an average age of 28.2 (SD = 9.2). 130 of the participants are male and 151 are female. The most popular countries of residence for participants are the US (27.5\%) and the UK (11.6\%). Of all other languages participants are fluent in, Spanish and French are the most popular.

\subsection{Agreement rates}

\begin{figure}[t]
	\centering
	\includegraphics[width=0.8\linewidth]{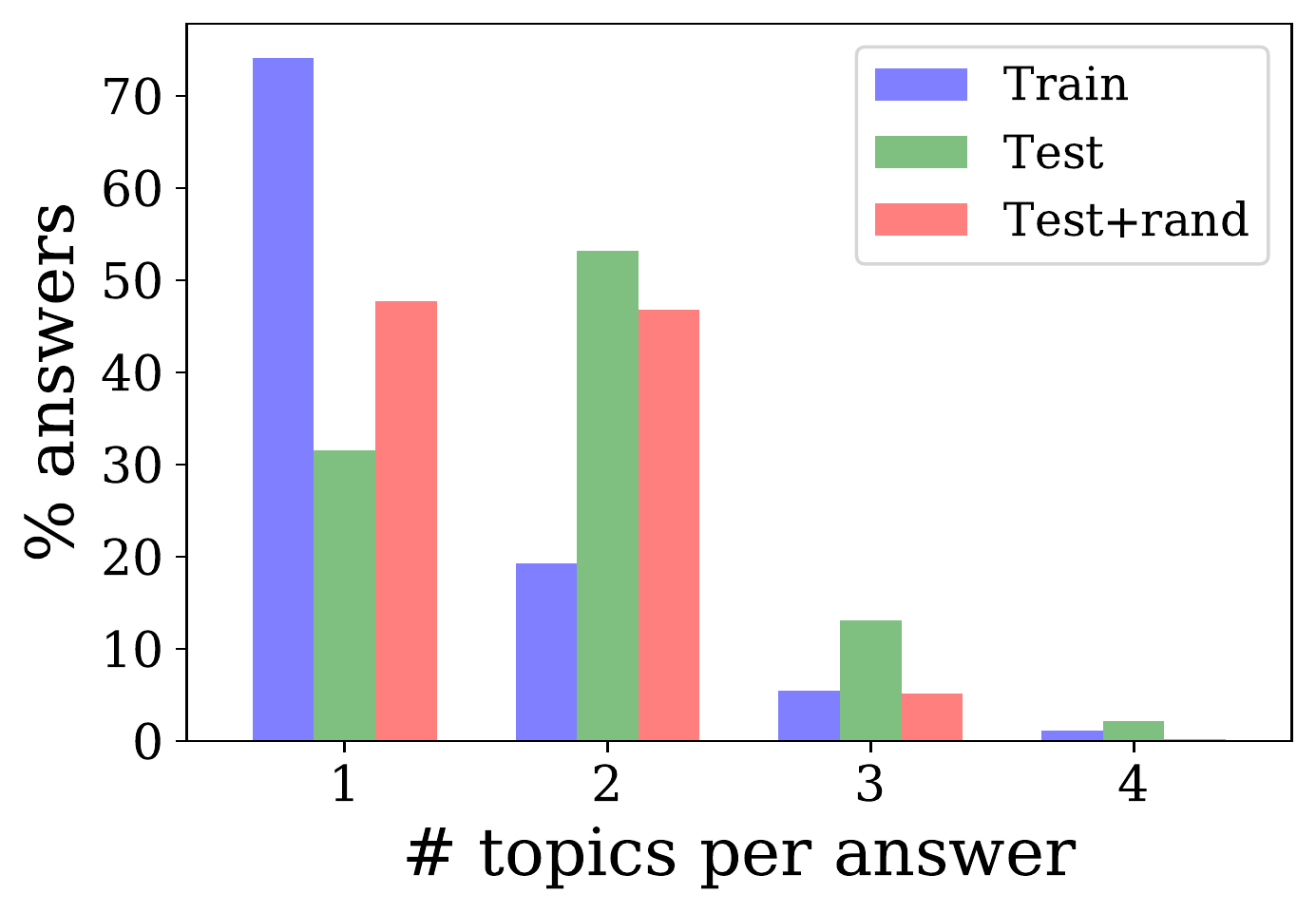}
	\caption{Numbers of topics chosen in each answer and their shares in \textit{train}, \textit{test} and \textit{test+rand} settings.}
	\label{fig:answer_lengths}
\end{figure}

After collecting responses, we had 3 answers per question. We report two types of agreement rates, described below.

\subsubsection{Post-level agreement rates.}

\begin{figure}[t]
	\centering
	\subfloat[]{%
		\includegraphics[width=1\linewidth]{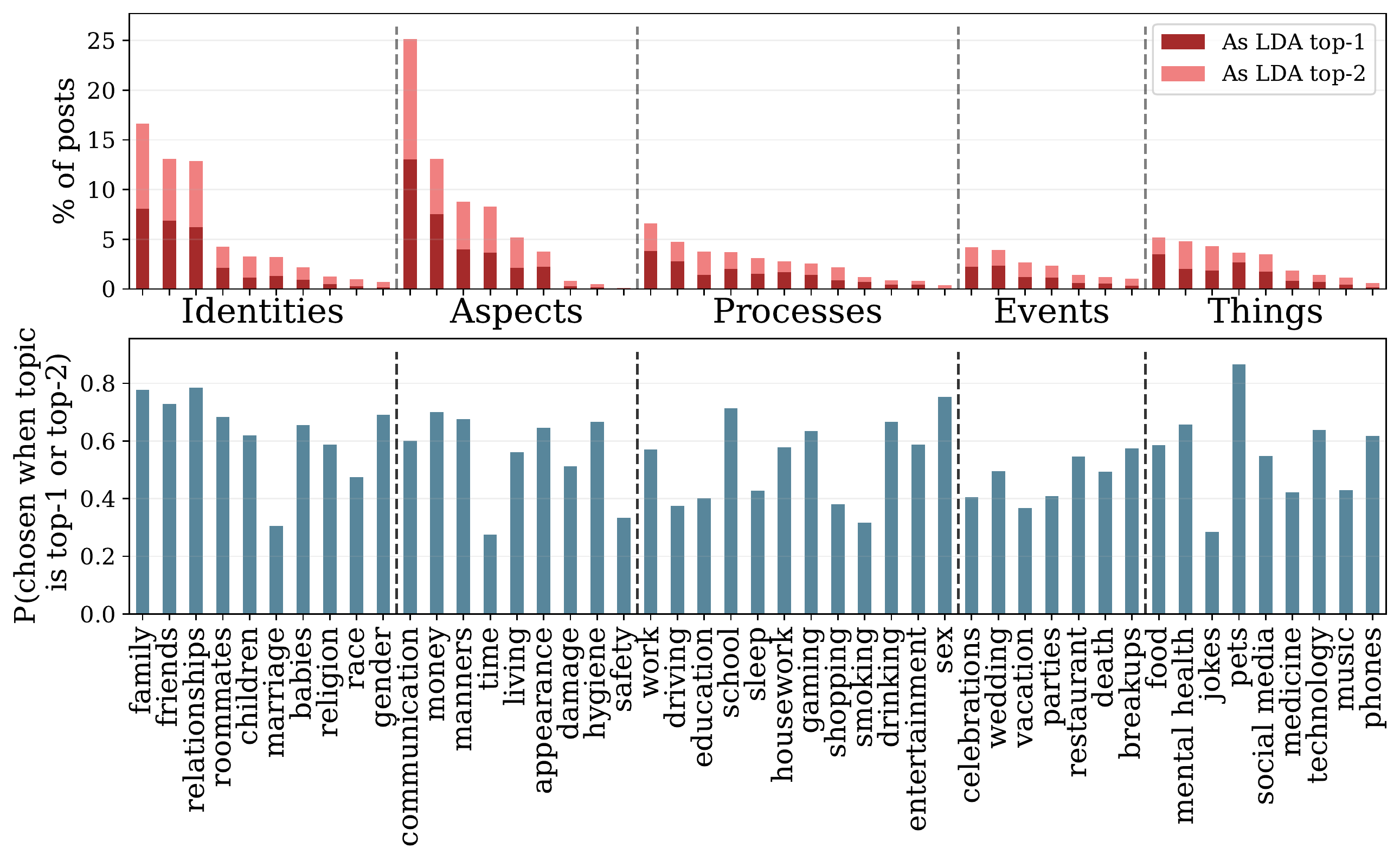}%
		\label{fig:prevalence_agreement_test}
	}\\
	\subfloat[]{%
		\includegraphics[width=1\linewidth]{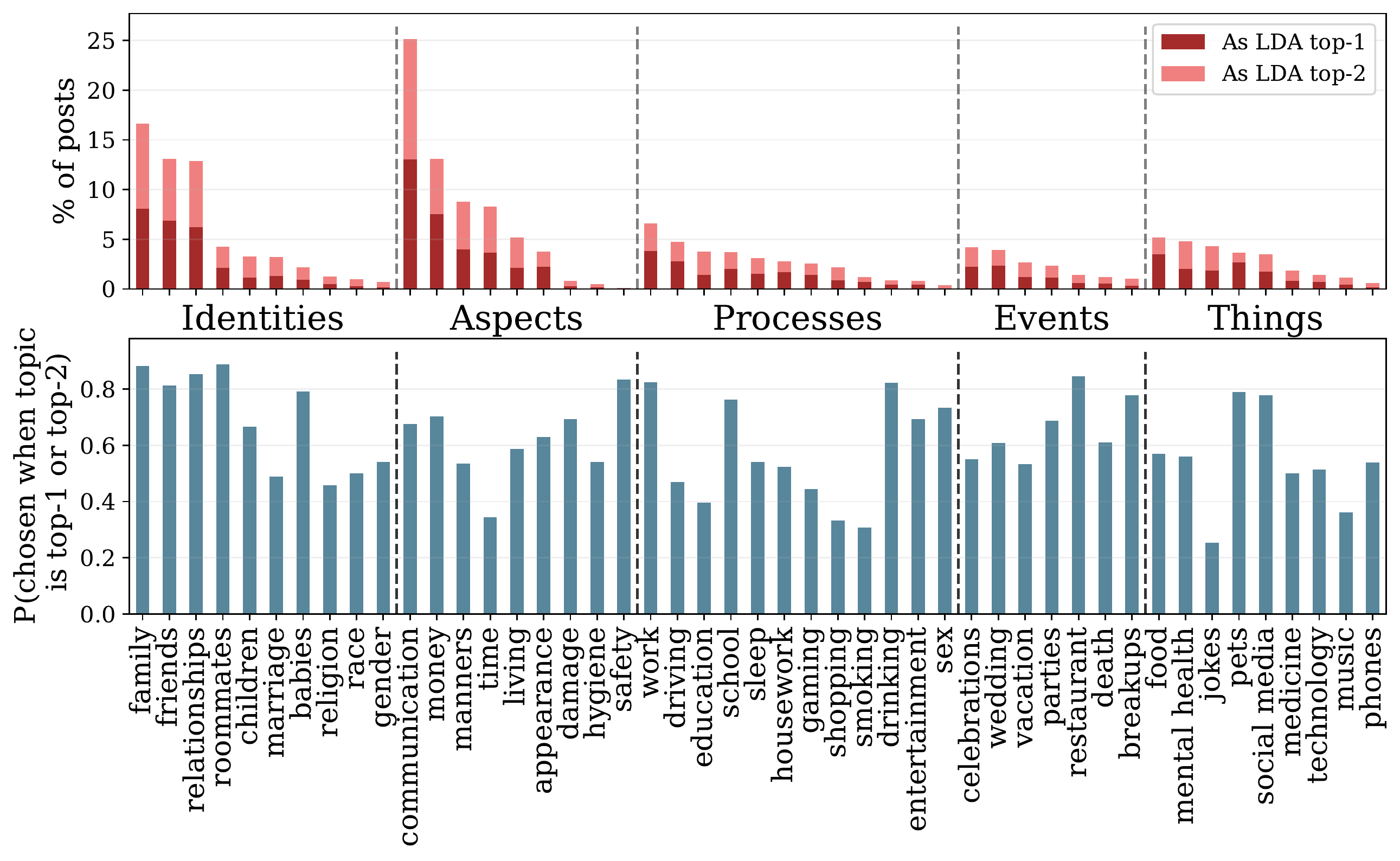}%
		\label{fig:prevalence_agreement_test_rand}
	}
	\caption{Prevalence (as a percentage of posts) and topic-specific agreement rate of topics in the \textit{test} set, comprising the first four months of 2020. \protect\subref{fig:prevalence_agreement_test} The agreement rate is from the \textbf{test} setting. \protect\subref{fig:prevalence_agreement_test_rand} The agreement rate is from the \textbf{test+rand} setting.}
	\label{fig:topic_stats_test}
\end{figure}

The quantities reported in Table 1 of the main paper are called post-level agreement rates, defined as the number of times an answer type appears in an answer, divided by the total number of answers for all questions. This rate is multiplied by $100$ in the table. For example, the agreement rate on the \textbf{top-1} row is the number of times the LDA top-1 topic appears in an answer, divided by the number of answers. So, for the \textbf{test} survey, which has $450 \times 3 = 1,350$ answers in total, 59.2\% of the answers contain the LDA top-1 topic. On the \textbf{top-1 or 2} row, this refers to the number of answers which contain either the LDA top-1 or top-2 topic, or both.

\subsubsection{Topic-specific agreement rate.}
The quantities reported in the lower bar plots of Fig. 4 (in the main paper) and each plot in \cref{fig:topic_stats_test} are called topic-specific agreement rates, which are defined for each topic. It is the number times topic $k$ appears in an answer, over the answers whose questions contain $k$ as either the LDA top-1 or top-2 topic. For example, suppose we wish to find the agreement rate for topic \topic{education}. There are $X$ number of questions with \topic{education} being the top-1 or top-2 topic, totaling $X \times 3$ answers recorded. Out of these $X \times 3$ answers, there are $Y$ answers which contain \topic{education}. The agreement rate is ${Y}/{(X \times 3)}$.

\section{Topic pairs}

In Section 5.2, we see that most dilemmas cover more than one topic. Particularly, we observe that the top two topics for each post do a much better job of describing the post's content, compared to its top-1 topic alone. In this section, we provide more information about topic pairs.

We focus on \emph{unordered} topic pairs. This means that a post with (top-1, top-2) topics of $(k, k')$ is in treated in the same group as a post with (top-1, top-2) topics of $(k', k)$. While the posterior probabilities indicate the order of salience between these top 2 topics, human experts find it difficult to find the correct order. In other words, posts are generally described by their two highest-scoring topics, but their order of salience is not easily recognized.

\subsection{Distribution of topic pair sizes}

Similar to individual topics, topic pairs vary significantly in size. We use the (log) complementary cumulative distribution (CCDF) function to display the distribution of these sizes. Out of $\binom{47}{2}$ unordered topic pairs, we find $33$ pairs ($3.1\%$) without a post, and only $259$ pairs ($24\%$) with at least 100 posts (Fig. 6 in the main paper). Finally, \cref{fig:topic_network} shows all named topics and the other topics they are most frequently associated with.

\subsection{Topic co-occurrence frequencies}

\begin{figure}
	\centering
	\includegraphics[width=1\linewidth]{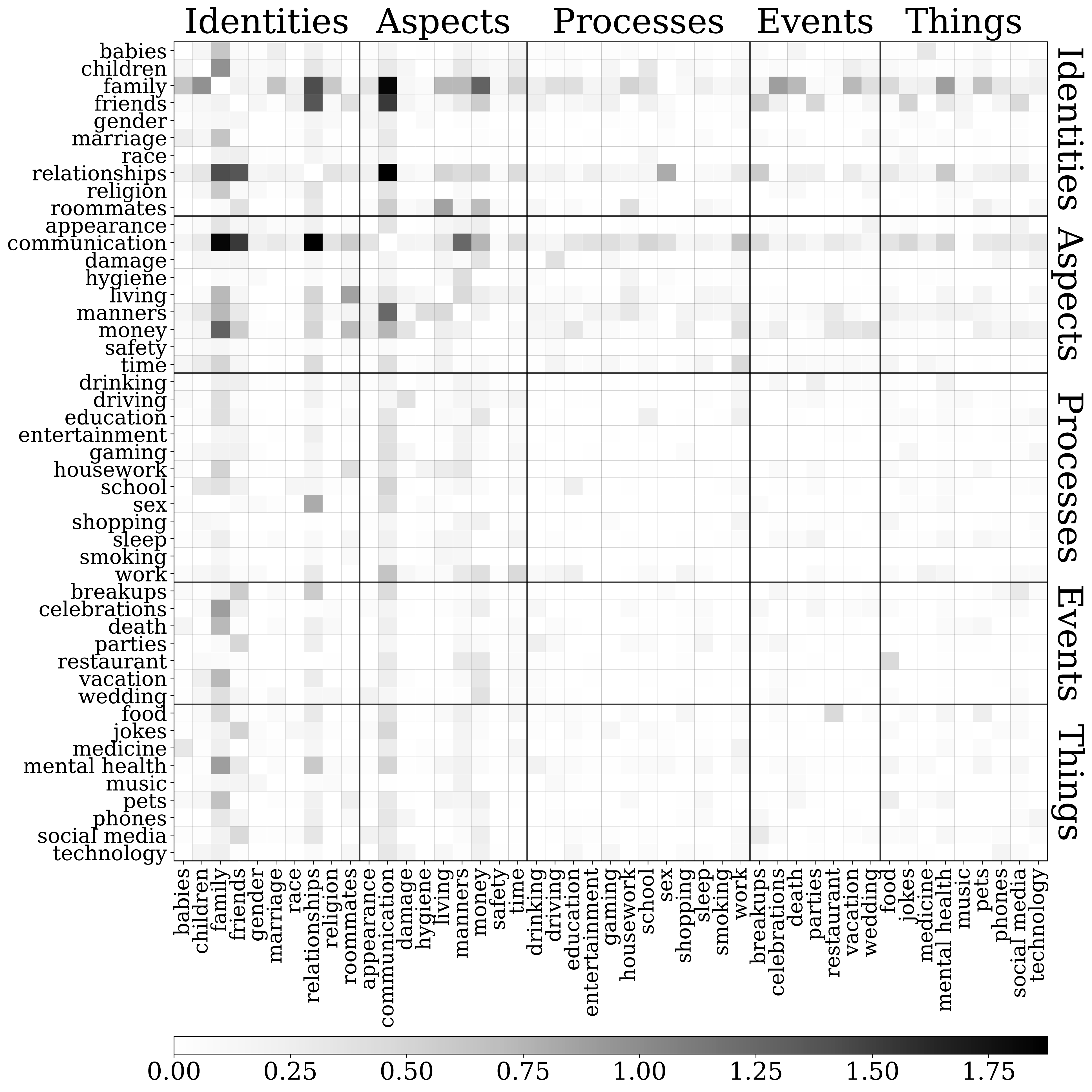}
	\caption{Topic co-occurrence in human answers. Each cell represents the frequency (as a percentage of all pairs) of a topic pair which appears in answers in the survey in Section 5. Rows and columns are organized into meta-categories.}
	\label{fig:pair_freq_human_annot}
\end{figure}

To assess how often a pair of topics $k$ and $k'$, we use two methods described below.

\subsubsection{Point-wise mutual information} 

Let $p(k)$ be the proportion of posts in the dataset whose top-1 topic is $k$, and $p(k, k')$ be the proportion of posts whose top-1 and top-2 topics are $k$ and $k'$, in either order of salience. If topics $k$ and $k'$ independently occur together, their joint distribution should be $p(k) p(k')$. To assess how more or less often $k$ and $k'$ co-occur than if they are independent, we compare their joint distribution $p(k, k')$ against the product of their marginals $p(k) p(k')$, in a metric called \emph{point-wise mutual information} (PMI) (Section 5.3):
\begin{align*}
	\text{PMI}(k, k') = \log_2 \frac{p(k, k')}{p(k) p(k')}.
\end{align*}
Fig. 7 in the main paper shows the PMI between all pairs of topics. Positive PMIs (red cells in the figure) correspond to pairs that are more likely to co-occur when assuming independence, whereas negative PMIs (blue cells) indicate less likely pairs.

\subsubsection{Topic co-occurrence in human answers}

The PMI is used to see topic co-occurrence within LDA. Using human input from the survey described in Section 5, we can measure which topic pairs tend to be chosen together by crowd-sourced workers.

Specifically, we look at answers containing at least two topics. For each answer, we extract every topic pair (so an answer of length $3$ gives $\binom{3}{2} = 3$ pairs). \cref{fig:pair_freq_human_annot} shows the frequency of each topic pair (as a percentage of all recorded pairs). As seen from the figure, most pairs chosen by human annotators are in the \emph{identities} and \emph{aspects} meta-categories. On the other hand, topics in \emph{processes}, \emph{events} and \emph{things} do not tend to co-occur with other topics in the same meta-category. Of all topics, \topic{communication}, \topic{family} and \topic{relationships} co-occur the most with other topics. This is expected, as these topics are relatively large in size.

\subsection{Voting and commenting patterns of topic pairs}
\label{appn:judgment_voting_stats}

\cref{fig:judgment_voting_stats} examines 
the judgement and voting statistics for topic pairs. We observe that the average post score (by voting) is positively correlated with the average number of comments (\cref{fig:judgment_voting_stats_a}). This is unsurprising in hindsight as both may be driven by the level of attention on a thread. \cref{fig:judgment_voting_stats_b} shows that smaller topic pairs have a wide spread of average topic scores, from around ten to several thousands, whereas large topics have scores around the global mean. \cref{fig:judgment_voting_stats_c} shows that smaller topic pairs have a wide spread of average post length, from $250$ to $450$ words, whereas large topics have lengths around $400$. We do not see a salient trend regarding the average \YA rate for topic pairs.

\subsection{Additional statistics on topics and topic pairs}

\Cref{table:LDA_clusters_full} lists the 70 LDA clusters, their mappings to the 47 named topics, their size (in number of posts) and coherence scores.

\Cref{fig:topic_network} visualises the (undirected) network of topics, by connecting two topics that has more than a 100 posts in the corresponding topic pair. 

{\onecolumn
	
	\begin{figure*}[t]
		\centering
		\subfloat[]{%
			\includegraphics[width=0.3\linewidth]{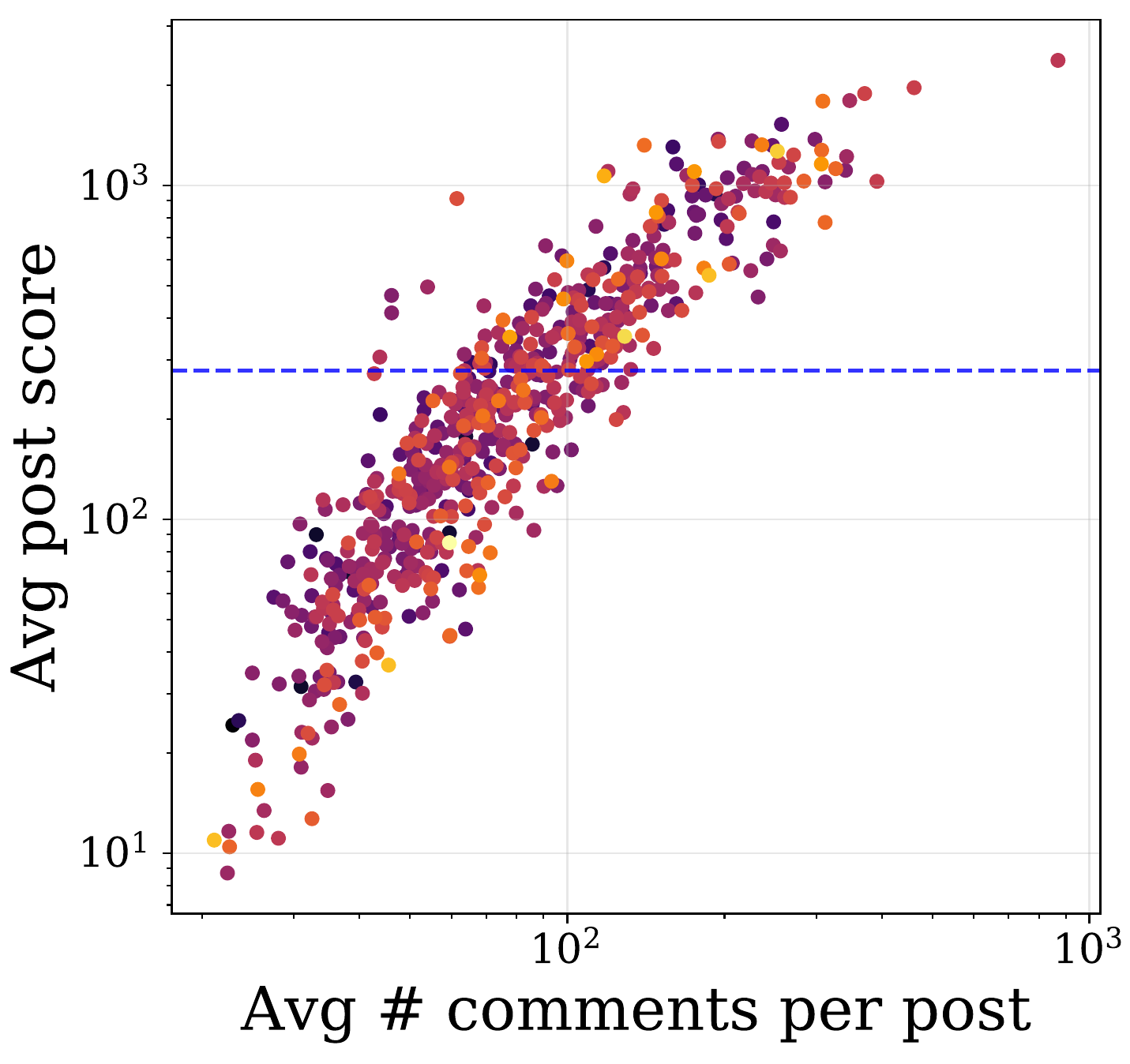}
			\label{fig:judgment_voting_stats_a}
		}
		\
		\subfloat[]{%
			\includegraphics[width=0.3\linewidth]{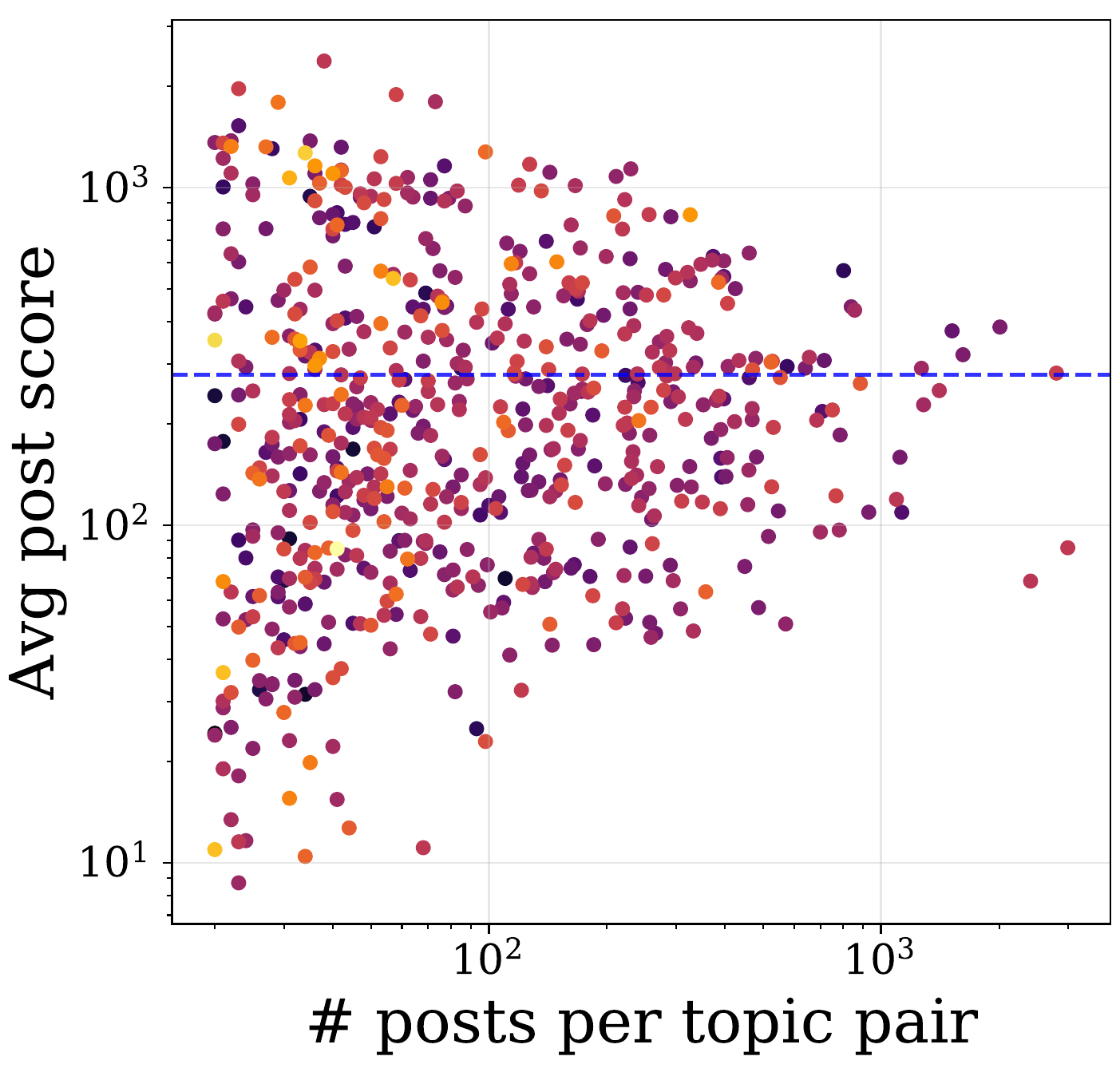}
			\label{fig:judgment_voting_stats_b}
		}
		\
		\subfloat[]{%
			\includegraphics[width=0.35\linewidth]{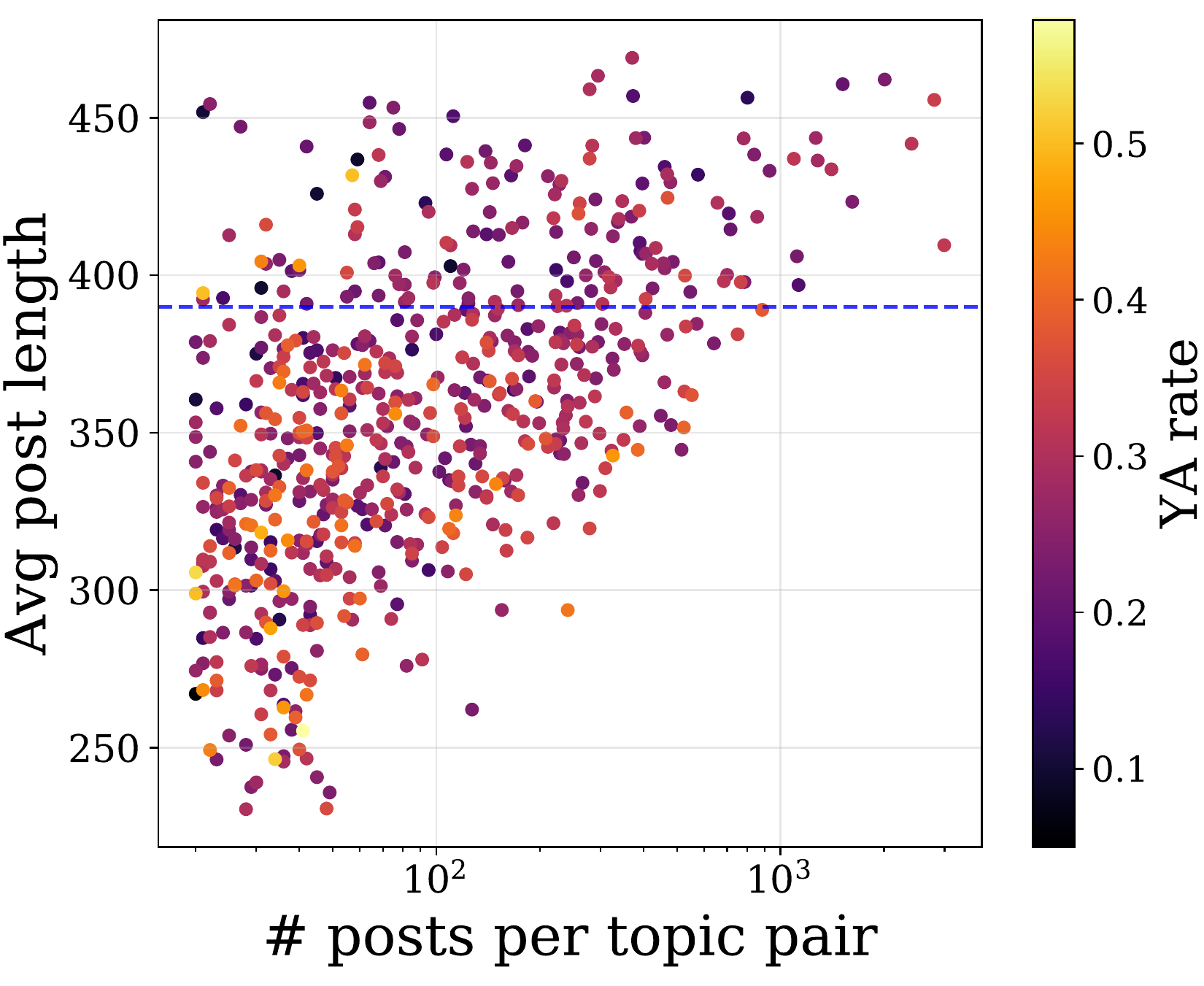}
			\label{fig:judgment_voting_stats_c}
		}
		\caption{Scatter plots for topic pairs. (a) Average number of comments per post versus average post score. (b) Number of posts versus average post score. (c) Number of posts versus average post length (in words). Topic pairs are colored by their \YA rate. A blue horizontal line indicates the $y$-axis mean over all posts.
		}
		\label{fig:judgment_voting_stats}
	\end{figure*}
	
	\begin{figure*}[t]
		\centering
		\includegraphics[width=0.8\linewidth]{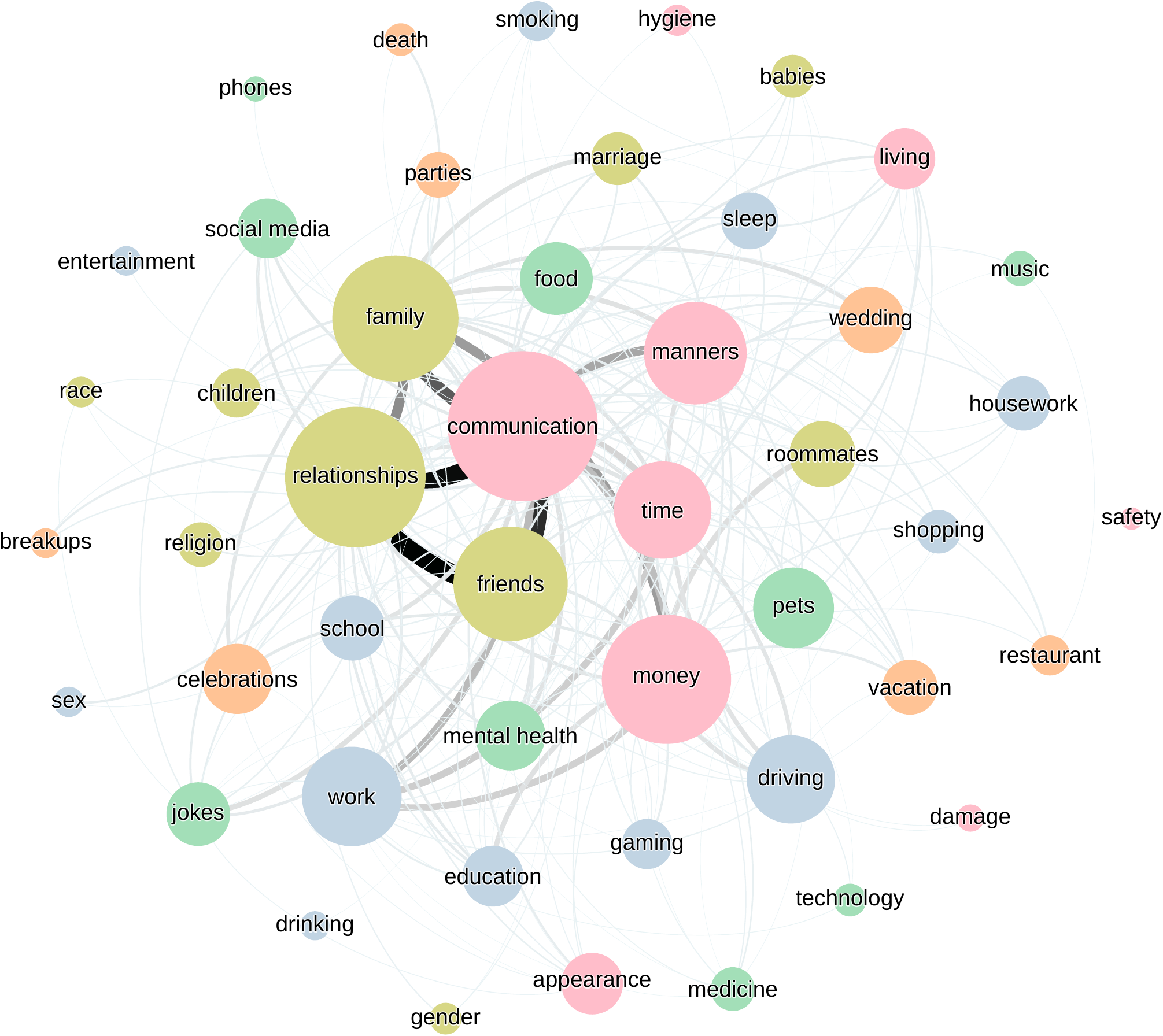}
		\caption{A network of co-occurring topics on \AITA. Topics are discovered and validated in Section 4.2. Node size denotes the number of posts and color represents its meta-category (yellow: \emph{identities}, pink: \emph{aspects}, blue: \emph{processes}, orange: \emph{events} and green: \emph{things}). Edge width is proportional to the number of posts in each topic pair. Only topic pairs with more than 100 posts are shown.}
		\label{fig:topic_network}
	\end{figure*}
	
	\clearpage
	
	\begin{longtable}{rlll}
		\caption{LDA clusters and their 10 most salient keywords. The topic names are from Section 4.2 of the main paper.}\\
		\toprule
		Topic name &         Size & Coherence &                                                                       Top-10 keywords \\
		\midrule
		appearance &        1,664 &     -2.74 &                     wear, hair, dress, look, like, shirt, clothe, shoe, makeup, color \\
		- &          137 &     -3.81 &                trash, tattoo, review, bug, garbage, bin, product, design, star, throw \\
		babies &          893 &     -1.42 &              baby, child, pregnant, kid, month, birth, pregnancy, week, husband, time \\
		breakups &          317 &     -1.66 &              ex, break, partner, year, relationship, month, cheat, date, contact, new \\
		celebrations &        2,484 &     -1.62 &                  birthday, gift, buy, like, year, present, thing, christma, day, card \\
		children &        1,257 &     -1.92 &                          kid, wife, son, child, old, year, young, boy, parent, nephew \\
		communication &        7,855 &     -1.16 &                      talk, like, try, thing, time, start, upset, way, apologize, come \\
		- &        1,448 &     -1.09 &                     text, message, send, texte, day, respond, reply, week, time, talk \\
		- &          448 &     -2.70 &    speak, language, english, customer, sorry, coffee, mistake, mobile, write, country \\
		- &           52 &     -3.32 &    jane, behavior, session, john, anna, discuss, kate, conversation, issue, therapist \\
		damage &          291 &     -3.16 &                  book, fix, damage, replace, pool, repair, read, new, paint, accident \\
		death &          508 &     -2.04 &      grandma, die, grandmother, pass, grandparent, family, mom, funeral, year, fiance \\
		drinking &          338 &     -2.49 &                   drink, bar, water, beer, bottle, alcohol, night, drunk, glass, wine \\
		driving &        1,494 &     -1.99 &                          car, drive, ride, pick, gas, driver, hour, way, minute, uber \\
		- &        1,047 &     -1.86 &                              road, light, turn, stop, pull, lane, way, walk, car, run \\
		- &          666 &     -2.05 &                park, spot, car, parking, street, lot, space, driveway, garage, people \\
		education &        1,829 &     -1.58 &            school, parent, year, college, high, graduate, job, live, university, work \\
		entertainment &          348 &     -2.56 &                 watch, movie, tv, coworker, film, office, worker, like, theater, work \\
		family &        1,328 &     -1.33 &                     sister, dad, mom, parent, year, old, family, sibling, live, young \\
		- &          830 &     -1.66 &         family, people, member, parent, like, holiday, come, time, year, thanksgiving \\
		- &          717 &     -2.57 &                        mum, drug, year, life, dad, mother, abuse, father, die, donate \\
		- &          478 &     -1.13 &                  mom, brother, parent, mother, old, young, come, little, family, live \\
		- &          232 &     -1.16 &                 daughter, cousin, aunt, uncle, family, old, year, parent, come, young \\
		food &        1,631 &     -2.53 &                           eat, food, lunch, like, bring, weight, buy, try, ice, pizza \\
		- &          635 &     -2.02 &                       cook, meal, dinner, meat, food, like, vegan, eat, chicken, dish \\
		friends &        5,959 &     -1.00 &                      friend, good, group, talk, hang, like, time, year, close, people \\
		gaming &        1,270 &     -2.24 &                  play, game, video, team, time, player, like, start, character, sport \\
		gender &          419 &     -2.57 &                   woman, man, gay, male, female, mark, come, people, gender, straight \\
		housework &        1,466 &     -2.18 &                clean, dish, leave, wash, laundry, thing, clothe, kitchen, mess, chore \\
		hygiene &          423 &     -3.23 &                 bathroom, shower, walk, smell, toilet, tooth, pee, nail, paper, brush \\
		jokes &        2,037 &     -1.66 &                     joke, like, people, laugh, comment, funny, thing, mean, fun, look \\
		living &          782 &     -1.88 &          door, open, building, window, close, knock, leave, come, apartment, neighbor \\
		- &          415 &     -1.99 &               room, living, space, bedroom, share, stuff, come, guest, small, kitchen \\
		- &          138 &     -2.03 &                     house, live, home, stay, buy, parent, rule, housemate, leave, let \\
		manners &        2,229 &     -1.58 &                      start, shit, yell, fuck, like, scream, fucking, try, throw, come \\
		- &        2,131 &     -2.45 &                            sit, seat, people, walk, look, lady, bus, gym, stand, wait \\
		marriage &        1,403 &     -2.01 &                husband, mother, father, year, child, law, divorce, marry, family, mil \\
		medicine &          921 &     -2.04 &       doctor, hospital, pain, sick, surgery, appointment, medical, need, help, cancer \\
		mental health &        2,132 &     -1.09 &                  issue, like, anxiety, mental, health, try, people, thing, time, help \\
		money &        4,505 &     -1.37 &                            pay, money, buy, month, job, work, help, need, spend, save \\
		- &        1,455 &     -1.99 &                 sign, pay, agree, charge, edit, check, state, situation, month, claim \\
		- &          528 &     -3.06 &                        ticket, sell, buy, card, concert, band, price, win, free, sale \\
		music &          592 &     -2.52 &                   music, tip, listen, song, play, loud, hear, server, like, headphone \\
		other &        6,694 &     -0.92 &                             time, try, thing, life, help, year, like, good, talk, bad \\
		- &           25 &     -4.10 &           bf, event, volunteer, bc, military, join, anime, army, organization, attend \\
		- &           16 &     -3.92 &               girlfriend, christma, secret, christmas, gf, year, eve, max, pen, santa \\
		parties &        1,071 &     -1.75 &              party, friend, night, drunk, leave, people, come, police, happen, invite \\
		pets &        1,947 &     -1.35 &                           dog, puppy, walk, time, bark, owner, let, like, care, leave \\
		- &          964 &     -2.00 &                       cat, pet, animal, adopt, care, kitten, shelter, vet, love, feed \\
		phones &          201 &     -3.42 &                   phone, number, jack, app, answer, look, check, cell, screen, iphone \\
		race &          424 &     -2.64 &                        guy, girl, white, dude, black, look, people, racist, race, boy \\
		relationships &        6,817 &     -1.06 &                    date, relationship, like, guy, meet, thing, talk, girl, love, year \\
		- &        1,661 &     -1.12 &                    boyfriend, time, come, stay, spend, visit, live, night, like, week \\
		- &           58 &     -4.46 &                club, dance, red, flag, camp, practice, beach, bff, strip, competition \\
		religion &          990 &     -2.60 &   church, believe, religious, people, opinion, religion, god, view, christian, belief \\
		restaurant &          859 &     -1.79 &                 order, restaurant, dinner, table, food, place, come, leave, pay, cake \\
		roommates &        2,263 &     -1.46 &                  roommate, live, apartment, rent, month, place, pay, lease, stay, new \\
		safety &           78 &     -3.91 &                         lock, steal, key, bike, door, milk, unlock, leave, locker, dh \\
		school &        2,063 &     -2.30 &             class, teacher, school, student, study, grade, test, project, exam, group \\
		sex &          346 &     -1.48 &                         sex, kiss, sexual, porn, time, condom, like, start, stop, try \\
		shopping &        1,023 &     -2.29 &                         store, bag, stuff, leave, pick, box, item, shop, grocery, buy \\
		sleep &        1,678 &     -1.36 &                     sleep, bed, night, wake, morning, asleep, couch, room, time, stay \\
		smoking &          796 &     -3.16 &  smoke, neighbor, weed, yard, cigarette, property, fence, smoking, tree, neighborhood \\
		social media &        1,778 &     -1.23 &            post, picture, photo, facebook, social, medium, people, send, like, friend \\
		technology &          490 &     -2.95 &          email, computer, laptop, report, internet, work, code, need, meeting, camera \\
		time &        4,570 &     -1.25 &                        work, day, time, hour, week, come, weekend, leave, night, plan \\
		vacation &        1,620 &     -1.47 &                    trip, vacation, plan, travel, hotel, day, time, week, year, flight \\
		wedding &        1,979 &     -1.45 &                wedding, invite, marry, plan, party, friend, attend, fiancé, year, day \\
		- &          213 &     -4.64 &     account, ring, throwaway, propose, mate, bet, fiancée, password, engagement, boat \\
		work &        4,642 &     -1.58 &                work, job, company, boss, manager, new, coworker, time, office, people \\
		\bottomrule
		\label{table:LDA_clusters_full}
	\end{longtable}
}

\section{Moral foundation word use for topics and topic pairs}

\begin{figure*}[t]
	\centering
	\includegraphics[height=1.2\linewidth]{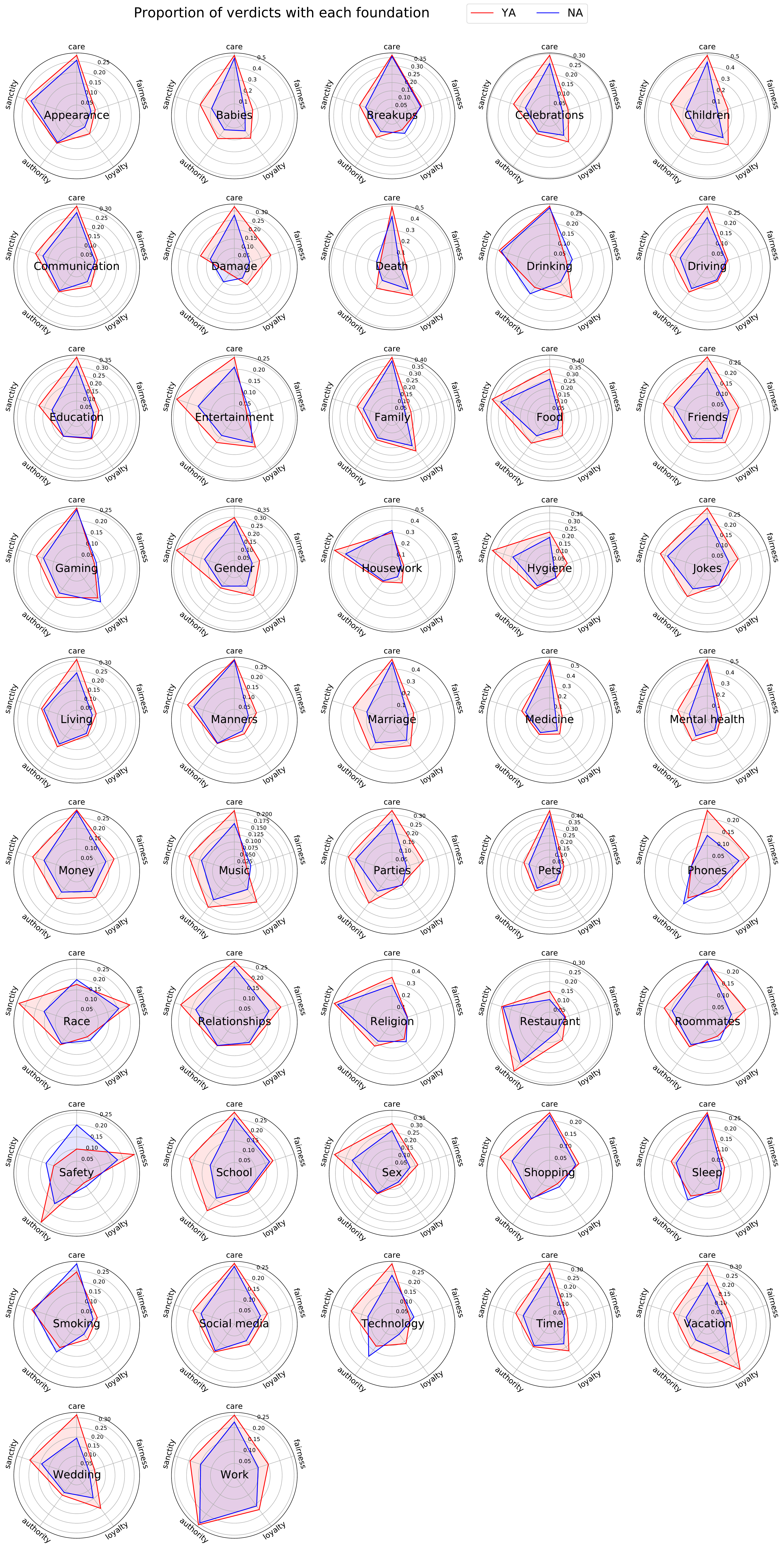}
	\caption{Strengths of moral foundations in each topic's verdict. In each radar plot, each pentagon's vertex represents the proportion of verdicts in a topic that have the presence of at least that foundation. Red pentagons represent \YA verdicts; blue pentagons represent \NA verdicts.}
	\label{fig:mfd_radar_comments}
\end{figure*}

\begin{figure*}[t]
	\centering
	\includegraphics[height=1.2\linewidth]{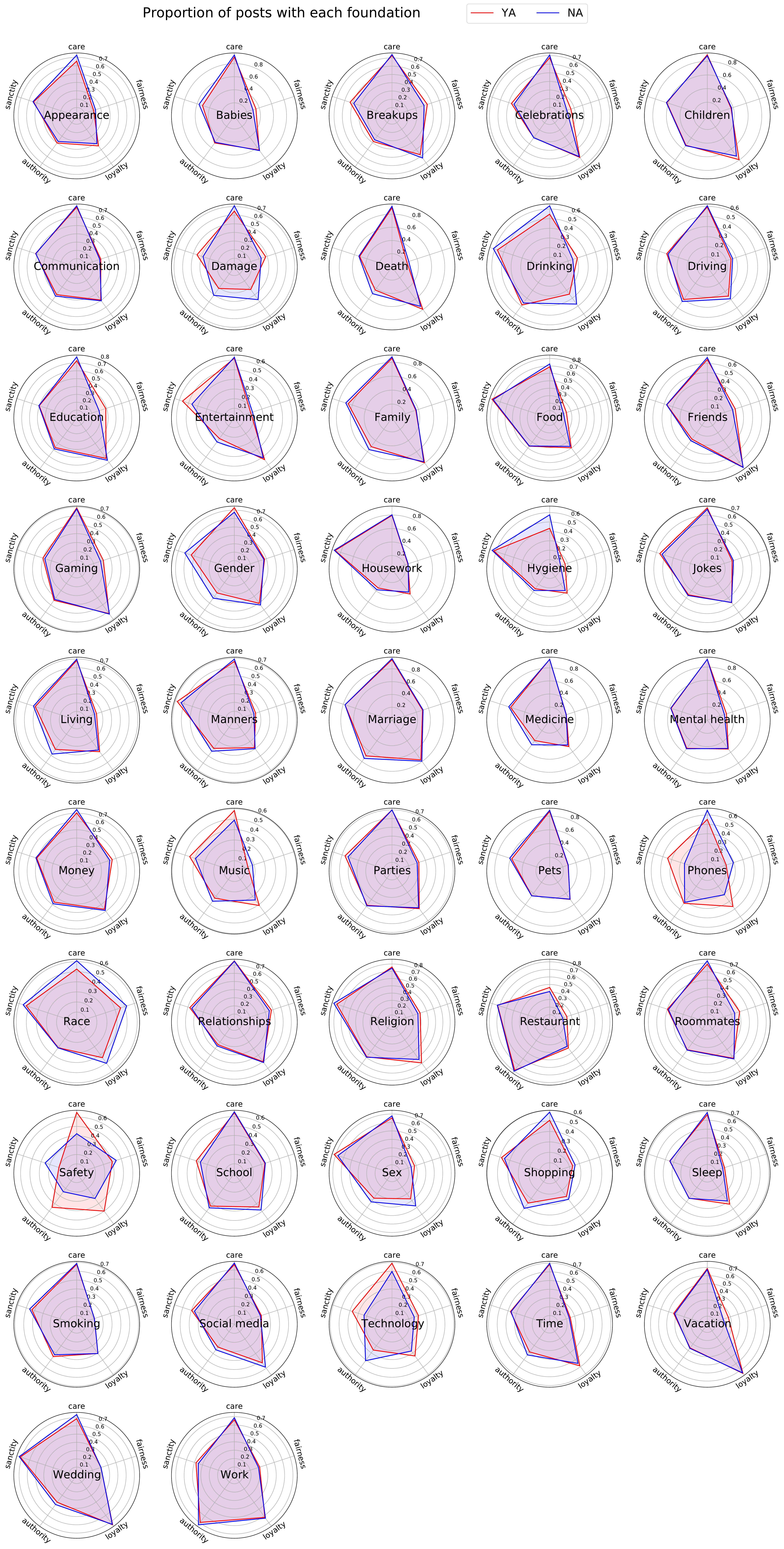}
	\caption{Strengths of moral foundations in each topic's post. In each radar plot, each pentagon's vertex represents the proportion of posts in a topic that have the presence of at least that foundation. Red pentagons represent \YA-judged posts; blue pentagons represent \NA-judged posts.}
	\label{fig:mfd_radar_posts}
\end{figure*}

\end{document}